\documentclass[prd,superscriptaddress]{revtex4}
\usepackage{amsmath}
\usepackage{amsfonts}
\usepackage{placeins}
\usepackage{latexsym}
\usepackage{subfigure}
\usepackage[colorlinks=true, linkcolor=blue, citecolor=red, urlcolor=magenta]{hyperref}
\usepackage{graphicx}
\usepackage{epsfig}
\usepackage{comment}

\def\eqref#1{{(\ref{#1})}}

\def\be{\begin{equation}}
\def\ee{\end{equation}}
\def\beq{\begin{eqnarray}}
\def\eeq{\end{eqnarray}}

\usepackage{orcidlink}

\begin{document}

\title{Noncommutative BTZ Gravastar in Gauss-Bonnet Gravity}

\author{M. A. Anacleto~\orcidlink{0000-0003-4625-7322}}
\email{anacleto@df.ufcg.edu.br}
\affiliation{Departamento de F\'{\i}sica, Universidade Federal de Campina Grande
Caixa Postal 10071, 58429-900 Campina Grande, Para\'{\i}ba, Brazil}
\affiliation{Unidade Acad\^emica de Matem\'atica, Universidade Federal de Campina Grande
\\
58429-900 Campina Grande, Para\'{\i}ba, Brazil}
\author{A. T. N. Silva~\orcidlink{0009-0007-0067-600X}}
\email{andersont.nsilva@gmail.com}
\affiliation{Departamento de F\'{\i}sica, Universidade Federal de Campina Grande
Caixa Postal 10071, 58429-900 Campina Grande, Para\'{\i}ba, Brazil}
\author{L. Casarini~\orcidlink{0000-0002-0869-9405}}
\email{lcasarini@academico.ufs.br}
\affiliation{Departamento de F\'isica, Universidade Federal de Sergipe
49100-000 Aracaju, Sergipe, Brazil}
\author{B. R. Carvalho~\orcidlink{0009-0002-2184-7588}}
\email{brunorego93@gmail.com}
\affiliation{Departamento de F\'{\i}sica, Universidade Federal de Campina Grande
Caixa Postal 10071, 58429-900 Campina Grande, Para\'{\i}ba, Brazil}

\pretolerance10000
\begin{abstract}
\noindent We constructed a (2+1)-dimensional gravastar whose interior region was characterized by a noncommutative geometry with Gauss-Bonnet corrections, while the exterior region was described by a Gauss-Bonnet-modified BTZ solution. The two geometries were joined across a thin shell using the generalized junction formalism, from which we derived the surface equations of state. We investigated the energy conditions at the shell, determining the parameter regimes under which the null, weak, and strong energy conditions were satisfied. We also analyzed the thermodynamic properties of the noncommutative interior region, including the temperature, entropy, and heat capacity.

\end{abstract}

\maketitle

\section{INTRODUCTION}\label{sec:introduction}

Recent advances in observational astrophysics have opened a window onto gravity in the strong-field regime. The horizon-scale imaging of M87$^{*}$~\cite{EventHorizonTelescope:2019dse,EventHorizonTelescope:2021bee,Banerjee:2019nnj} and Sgr~A$^{*}$~\cite{EventHorizonTelescope:2022wkp,Banerjee:2022jog}, the observation of electromagnetic emission from accreting compact objects~\cite{Czerny:2022xfj}, and the detection of gravitational waves from compact-binary mergers by LIGO/Virgo~\cite{KAGRA:2021duu} have provided independent and complementary channels for probing the geometry of spacetime near highly compact sources. General relativity has so far provided a successful framework for interpreting these phenomena. At the same time, the extrapolation of its classical solutions to the most extreme regimes leads to unresolved issues, most notably the appearance of spacetime singularities and the black-hole information-loss problem. Understanding whether these features signal limitations of the classical description, or whether they can be resolved through modifications of the underlying geometry, remains an important question in gravitational physics. This motivates the study of modified black-hole geometries and horizonless compact configurations as theoretical laboratories in which deviations from the standard picture can be systematically investigated.

In this context, the gravastar (gravitational vacuum star) scenario proposed by Mazur and Mottola~\cite{Mazur:2001fv,Mazur:2004fk} provides a particularly intriguing framework for constructing compact objects that avoid the central singularity present in classical black-hole solutions. In its original formulation, the configuration consists of three regions characterized by distinct equations of state. A de Sitter core satisfying $p=-\rho$ occupies the interior, while a shell of stiff matter obeying $p=\rho$ separates the two geometries. The exterior vacuum region, where $p=\rho=0$, is described by the Schwarzschild spacetime~\cite{Schwarzschild:1916uq}. Replacing the conventional black-hole interior with a regular vacuum core provides a simple mechanism for constructing highly compact configurations while simultaneously avoiding the formation of a central singularity. Since its introduction, the gravastar scenario has been investigated from different perspectives, including stability, tidal properties, gravitational-wave signatures, gravitational lensing, and shadow phenomenology~\cite{Chirenti:2007mk,Pani:2009hk,Uchikata:2016qku,Kubo:2016ada,Pani:2015tga,Sakai:2014pga,Pani:2012zz,MartinMoruno:2011rm,Pani:2010em,Pani:2009ss}.

Noncommutative-geometry-inspired models provide another approach to regularizing compact-object geometries by introducing a minimum length scale $\sqrt{\theta}$. Nicolini et al.~\cite{Nicolini:2005vd} were among the first to incorporate spacetime noncommutativity into black-hole solutions through a smeared matter distribution modeled by a Gaussian profile. Since then, noncommutative geometries have been extensively investigated using different matter distributions and geometric constructions~\cite{Anacleto:2019tdj,Zeng:2021dlj,Zeng:2022fdm,Campos:2021sff,Anacleto:2020efy,Anacleto:2020zfh,Hu:2023eow,Anacleto:2022shk,Saleem:2023pyx,AraujoFilho:2024rss,Hamil:2024ppj,Wang:2024fiz,Jha:2022bpv,Jha:2023htn}. Within this broader framework, Lobo and Garattini~\cite{Lobo:2010uc} examined gravastar solutions in noncommutative geometry and investigated their physical properties. \"Ovg\"un et al.~\cite{Ovgun:2017jzt} constructed a four-dimensional charged thin-shell gravastar in noncommutative geometry and investigated its stability. More recently, Anacleto, Silva, and Casarini~\cite{Anacleto:2026hkb} investigated a four-dimensional noncommutative thin-shell gravastar, examining its stability and optical properties through photon trajectories and gravitational deflection. In the BTZ setting, a Lorentzian matter distribution was used to incorporate noncommutative effects, which were shown to support the corresponding gravastar configuration even in the absence of a cosmological constant~\cite{Silva:2023bsu}. A subsequent study
~\cite{Anacleto:2025oef} investigated thin-shell gravastars in BTZ geometries with a minimum length,
examining their stability and thermodynamic properties.

A further relevant direction in the search for physically viable compact objects beyond general relativity involves higher-curvature corrections. The Gauss-Bonnet combination of quadratic curvature invariants is distinguished by preserving second-order metric field equations in dimensions where it contributes dynamically. In three dimensions, however, the ordinary Gauss-Bonnet density vanishes identically. Nontrivial lower-dimensional formulations can instead be obtained through carefully defined limiting procedures that introduce additional scalar interactions~\cite{Hennigar:2020fkv}. These formulations provide a setting in which corrections inherited from higher-dimensional gravity can affect the geometry and physical properties of compact configurations. Exploring gravastar models within such Gauss-Bonnet-inspired frameworks can therefore offer insight into the role of these corrections in horizonless compact configurations.

Lower-dimensional gravitational models have proven especially useful for probing both the classical and quantum aspects of gravity. In particular, the BTZ black hole, a $(2+1)$-dimensional solution with a negative cosmological constant proposed by Ba\~nados, Teitelboim, and Zanelli~\cite{Banados:1992wn}, provides an analytically tractable setting in which many properties of black-hole physics can be studied in detail without the technical complications of higher dimensions. Because of this combination of mathematical simplicity and physical richness, the BTZ geometry has become a valuable testbed for modifications of general relativity and their consequences for compact objects. Recent constructions of BTZ-like solutions incorporating Gauss-Bonnet contributions~\cite{Hennigar:2020fkv,Cuadros-Melgar:2009hih,Hennigar:2020drx} have further enriched this setting, providing a natural framework in which to examine how modifications of the underlying geometry affect compact gravitational configurations.

Motivated by these developments, in this work we construct a $(2+1)$-dimensional thin-shell gravastar by combining noncommutative geometry with Gauss-Bonnet corrections in a BTZ background. We first investigate the noncommutative Gauss-Bonnet BTZ geometry and its thermodynamic properties, including the horizon structure, Hawking temperature, entropy, and specific heat capacity, in order to characterize the effects of the two deformation parameters. We then construct the gravastar by matching a noncommutative Gauss-Bonnet interior to a Gauss-Bonnet modified BTZ exterior across a thin shell and derive the corresponding surface quantities through the junction formalism. The energy conditions at the shell are subsequently analyzed as functions of the noncommutative parameter, the Gauss-Bonnet coupling, and the cosmological constant. Finally, to distinguish the role played by the higher-curvature correction on each side of the junction, we separately examine configurations in which the Gauss-Bonnet contribution is retained only in the exterior or only in the interior region.

The remainder of this paper is organized as follows. In Sec.~\ref{sec:geometry-thermodynamics}, we introduce the noncommutative Gauss-Bonnet BTZ geometry and investigate its horizon and thermodynamic properties. In Sec.~\ref{sec2}, we construct the thin-shell gravastar, derive the surface energy density and pressure, and analyze the corresponding energy conditions, including the cases in which the Gauss-Bonnet correction is restricted to either the exterior or the interior geometry (Subsecs.~\ref{subsec:gb-exterior} and~\ref{subsec:gb-interior}, respectively). Finally, Sec.~\ref{con} summarizes our main results and conclusions.

\section{NONCOMMUTATIVE GAUSS-BONNET BTZ GEOMETRY AND THERMODYNAMICS}\label{sec:geometry-thermodynamics}

In this section, we introduce the noncommutative Gauss-Bonnet BTZ geometry used to characterize the interior region of the gravastar and investigate its thermodynamic properties. We consider a neutral three-dimensional Gauss-Bonnet BTZ geometry in which noncommutative effects are incorporated through a radially symmetric Lorentzian mass distribution. We then analyze how the noncommutative parameter and the Gauss-Bonnet coupling modify the horizon structure, Hawking temperature, entropy, and heat capacity.

The noncommutative contribution is introduced through the Lorentzian mass density~\cite{Liang:2012vx,Anacleto:2020efy},
\begin{eqnarray}
\rho_{\theta} = \frac{M_0 \sqrt{\theta}}{2 \pi (r^{2} + \theta)^{3/2}}. 	
\end{eqnarray} 
Here, $\theta$ is the noncommutative parameter with dimensions of length squared.  

The static line element is expressed as
\begin{eqnarray}
ds^{2} &=& -f_{GB}(r)\,dt^{2}
+\frac{dr^{2}}{f_{GB}(r)}
+r^{2}d\phi^{2},
\end{eqnarray}
where the metric function is given by~\cite{Hennigar:2020fkv}
\begin{eqnarray}
f_{GB}(r)
&=&
-\frac{r^{2}}{2\alpha} \left[ 1-\sqrt{ 1+\frac{4\alpha}{r^{2}}
\left(\frac{r^2}{l^{2}} - \mathcal{M}_{\theta}\right)
}\right],
\\
&=&
- \mathcal{M_{\theta}} + \frac{r^{2}}{l^2} + \frac{2\mathcal{M_{\theta}}~\alpha}{l^2} - \frac{\mathcal{M_{\theta}}^2~\alpha}{r^2} - \frac{r^2\alpha}{l^4} + \mathcal{O}(\alpha^2).
\end{eqnarray}
Here $l$ is the AdS radius related to the cosmological constant by $\Lambda = - 1/l^2$, $\alpha$ is the Gauss-Bonnet coupling constant and $\mathcal{M_{\theta}}$ is given by
\begin{eqnarray}
\mathcal{M_{\theta}} = \int_{0}^{r} \rho(r)2 \pi r dr = M_0\left(1 - \frac{\sqrt{\theta}}{\sqrt{r^{2} + \theta}} \right). 	
\end{eqnarray} 
For $\alpha\ll 1$ and $\theta \ll1$, we can write the metric function as follows
\begin{eqnarray}
f_{GB}(r)
&=&
- M_0\left(1 - \frac{\sqrt{\theta}}{\sqrt{r^2 + \theta}} \right)\left(1 - \frac{2\alpha}{l^2}\right) + \frac{r^{2}}{l^2} \left(1 - \frac{\alpha}{l^2}\right) - \frac{\alpha M^2_0}{r^2}\left(1 - \frac{2\sqrt{\theta}}{\sqrt{r^2 + \theta}}\right) ,
\\
&=&
- M_0\left(1 - \frac{2\alpha}{l^2} - \frac{\sqrt{\theta}}{r} \right) + \frac{r^{2}}{l^2} \left(1 - \frac{\alpha}{l^2}\right)- \frac{\alpha M^2_0}{r^2}.
\end{eqnarray}
For $\alpha=0$, the metric function of the noncommutative BTZ black hole is recovered~\cite{Anacleto:2020efy}.

The event horizon is determined by the condition $f_{GB}(r) = 0$
\begin{eqnarray}
 r^{4} - M_0 l^2\left(1 - \frac{\alpha}{l^2}\right)r^{2} + \sqrt{\theta}~l^2 M_0r - \alpha l^2M^2_0 = 0, 
\end{eqnarray}
or, equivalently~\cite{Anacleto:2020efy},
\begin{equation}
\tilde{r}_h
\approx
r_h
\left(
1+\frac{\alpha}{2l^{2}}
\right)
-
\frac{\sqrt{\theta}}{2}, \quad \mbox{or} 
\quad 
\tilde{M}\approx \frac{r^2_h}{l^2}\left(1 + \dfrac{\alpha}{l^2}-\frac{\sqrt{\theta}}{r_h}\right)
= M_0\left(1+ \dfrac{\alpha}{l^2}- \sqrt{\frac{\theta}{l^2 M_0}}\right).
\label{eq:rplus-compact}
\end{equation}
Eq.~\eqref{eq:rplus-compact} explicitly displays the leading
corrections induced by noncommutative geometry and by the
Gauss--Bonnet coupling. In the limit $\alpha\rightarrow 0$, the noncommutative BTZ result is recovered, $\tilde{r}_h \simeq r_h - {\sqrt{\theta}}/{2}$~\cite{Anacleto:2020efy}.

For the Hawking temperature, we have
\begin{eqnarray}
\tilde{T}_H=\frac{\tilde{r}_h}{2\pi l^2}\left(1 - \frac{\alpha}{l^2}\right) + \frac{\alpha M_0^2}{2 \pi \tilde{r_h}^3} - \frac{M_0\sqrt{\theta}}{4\pi \tilde{r}^2_h}.   
\end{eqnarray}
In terms of $r_h$, we obtain
\begin{eqnarray}
\tilde{T}_H=\frac{M_0}{2\pi {r}_h} \left( 1 - \frac{\alpha M_0}{2 r_h^2} - \frac{\sqrt{\theta}}{r_h}\right).   
\end{eqnarray}
The result can be expressed in Schwarzschild-type form as follows~\cite{Silva:2023bsu,Anacleto:2025oef}:
\begin{eqnarray}
\mathcal{T}_H=\frac{\tilde{T}_H}{M_0}=\frac{1}{2\pi r_h}\left(1 - \frac{\tilde{\alpha}}{2r_h^2} - \frac{\sqrt{\theta}}{r_h}\right),   
\end{eqnarray}
where, $\tilde{\alpha} = \alpha M_0$.
\begin{figure}[htbh!]
\centering
\subfigure[]{\includegraphics[scale=0.22]{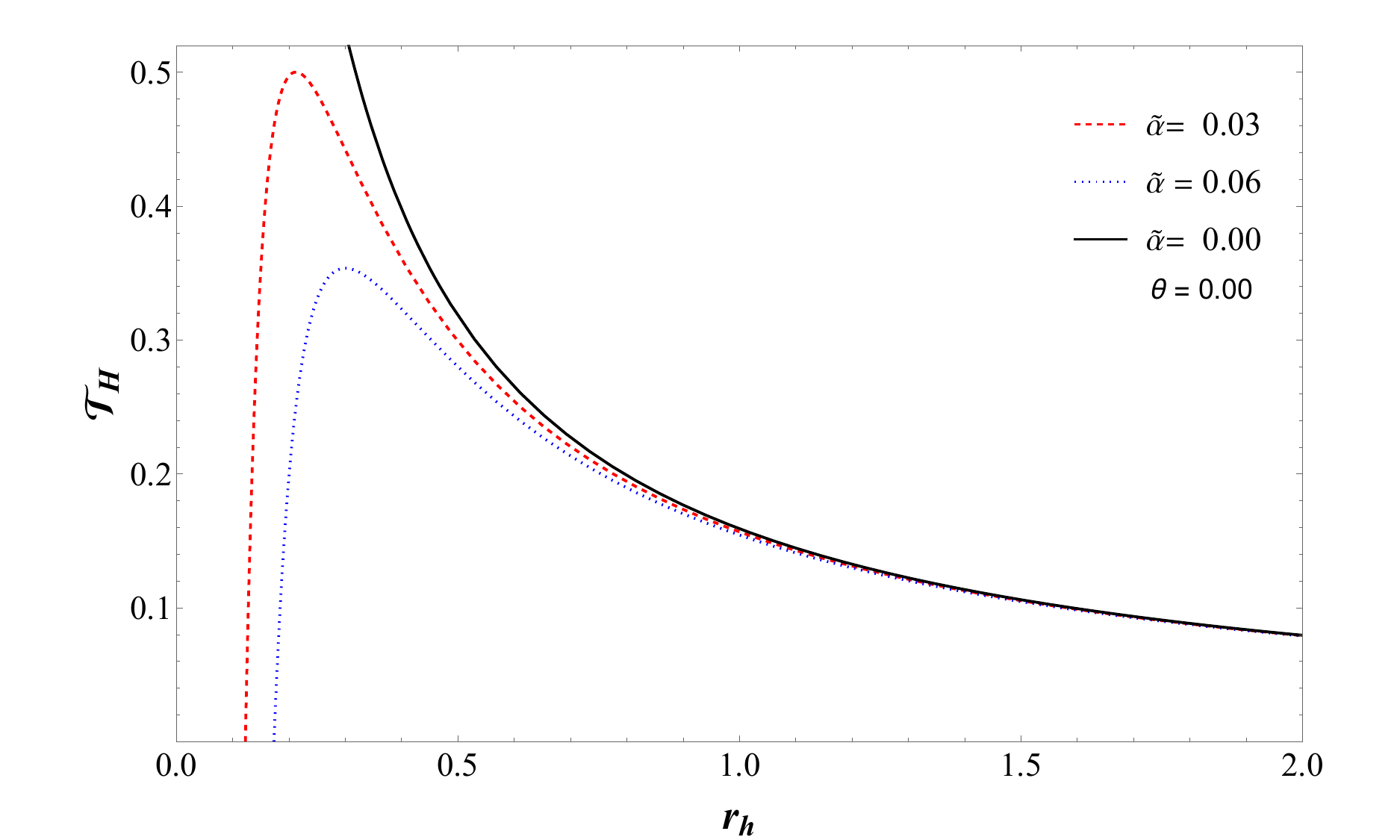}}
\subfigure[]{\includegraphics[scale=0.22]{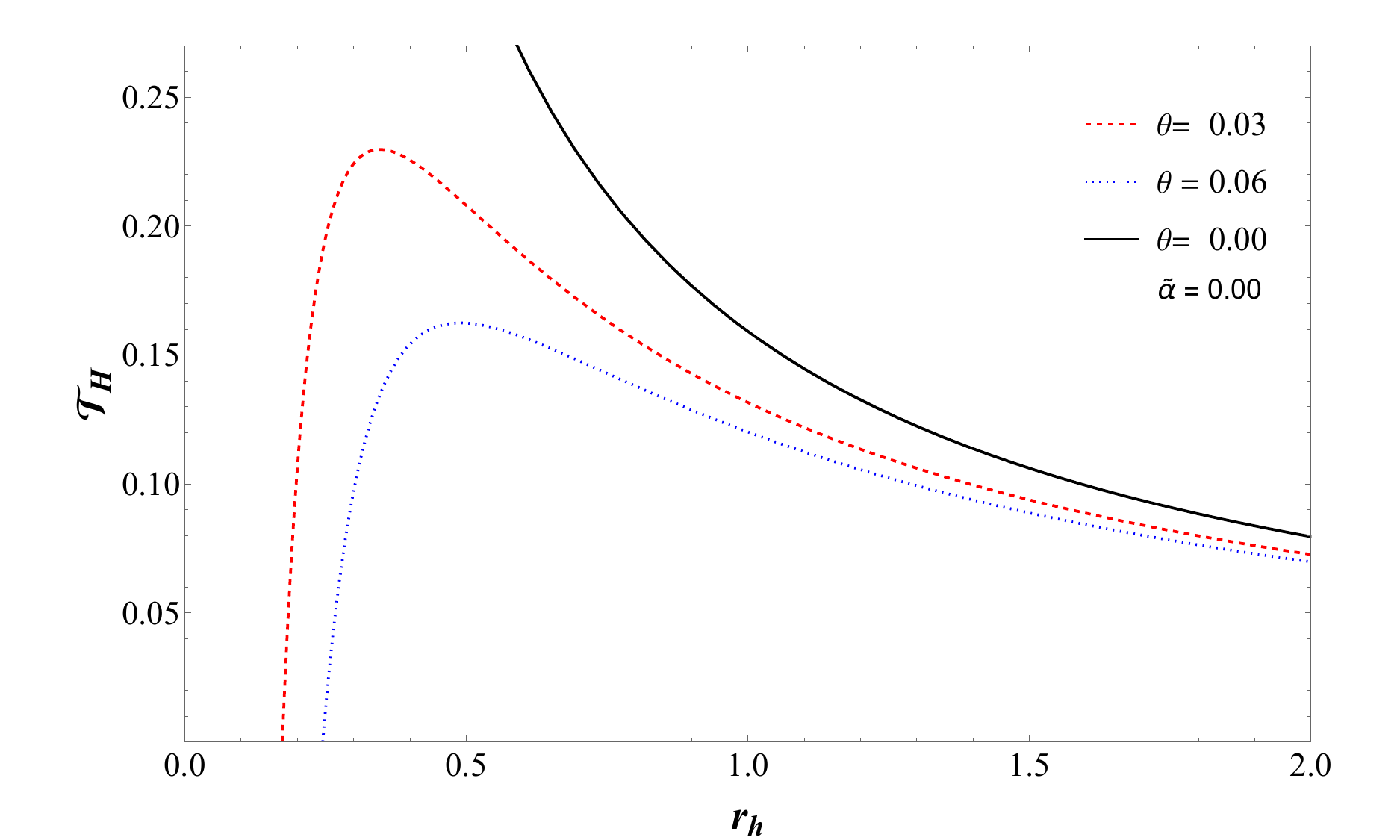}}
 \caption{Hawking temperature $\mathcal{T}_H$ as a function of horizon radius $r_h$. (a) Gauss-Bonnet coupling for various values of $\tilde{\alpha}$ with $\theta=0$. (b) Noncommutative parameter for different values of $\theta$ with $\tilde{\alpha}=0$. In both scenarios, the introduced corrections regularize the behavior of the Hawking temperature in the small-radius region, establishing a finite maximum temperature, whereas it would diverge in the absence of these corrections.}
\label{fig:temperature_individual_effects}
\end{figure}
\begin{figure}[htbh!]
\centering
\subfigure[]{\includegraphics[scale=0.22]{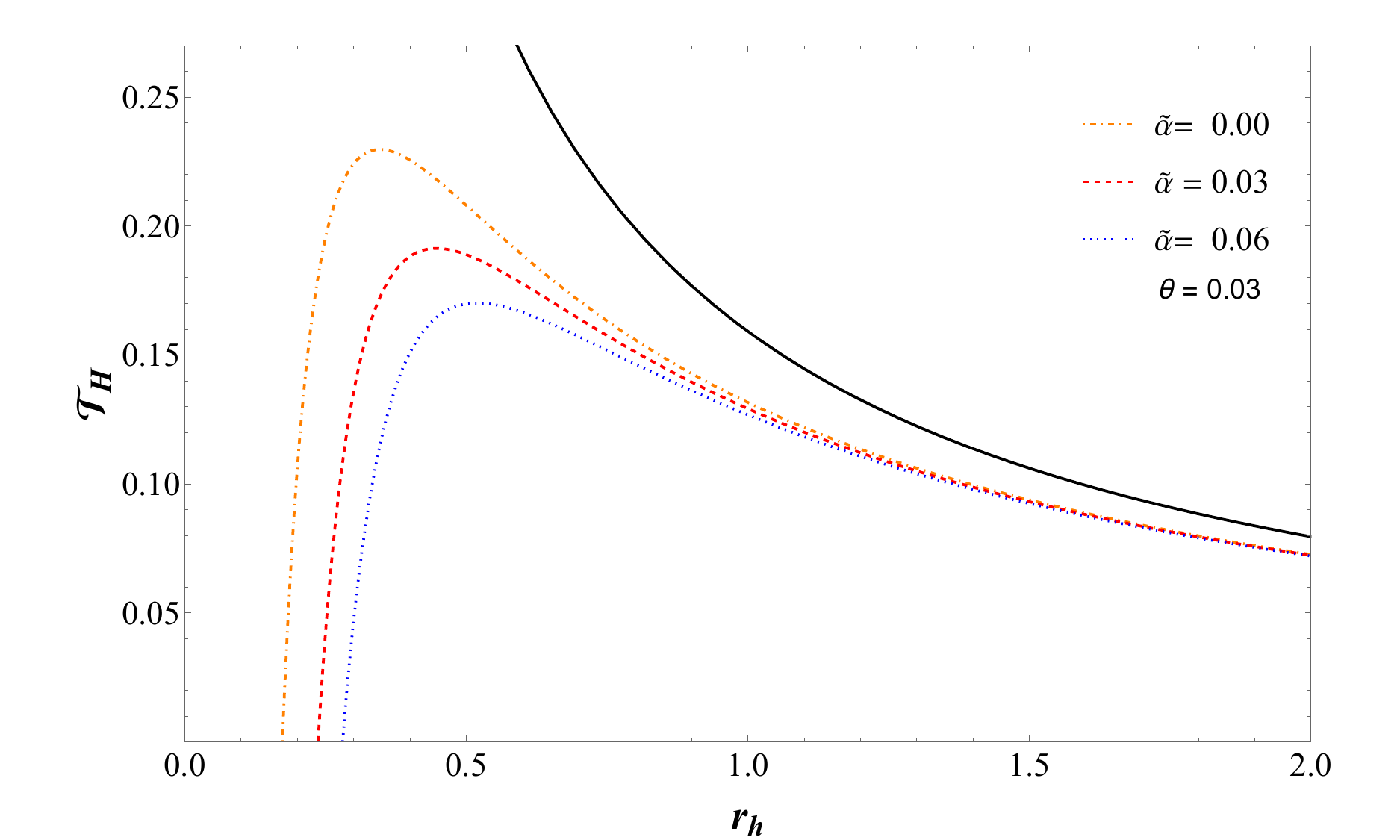}}
\subfigure[]{\includegraphics[scale=0.22]{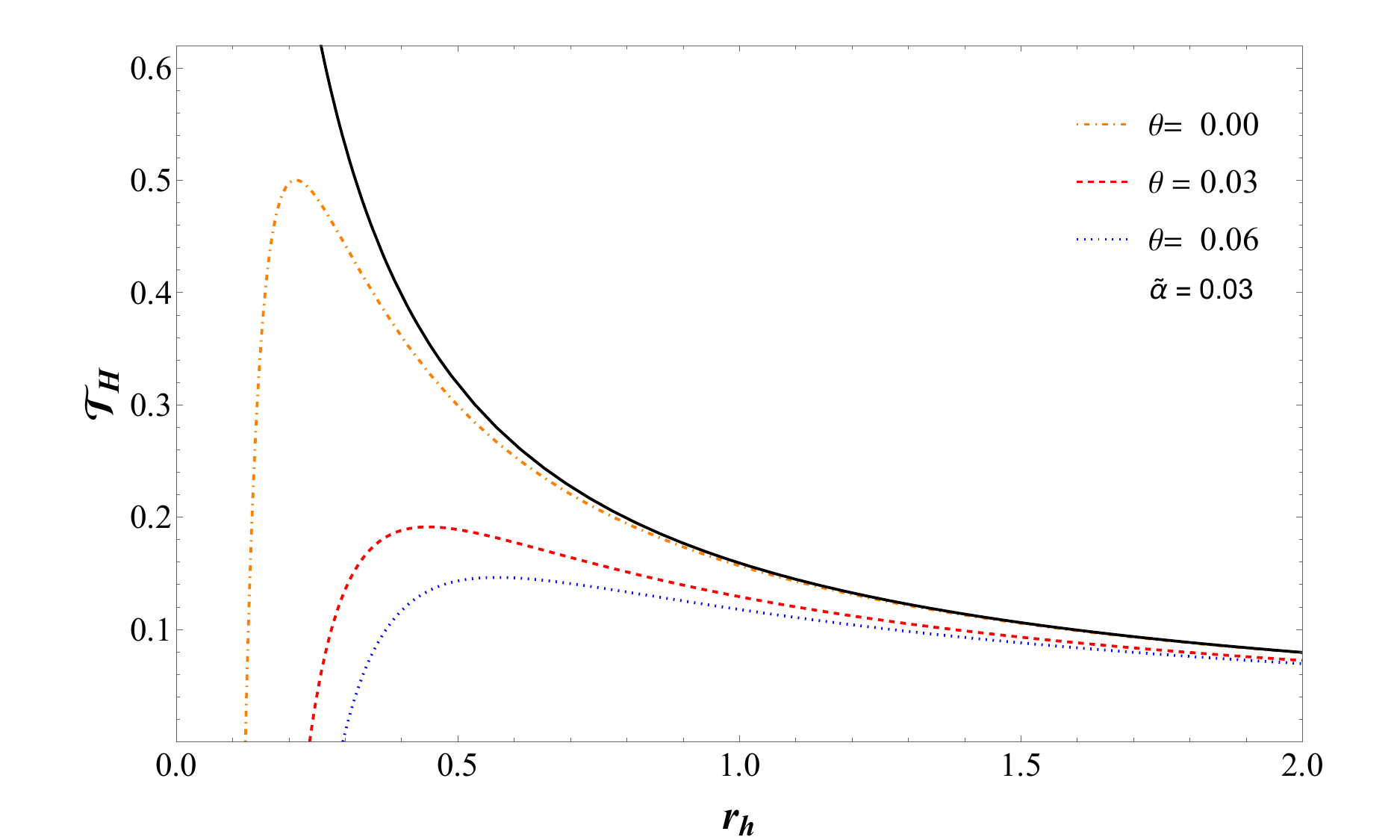}}
 \caption{Combined effects of Gauss-Bonnet and noncommutative corrections on the Hawking temperature. (a) \(\mathcal{T}_H\) for different values of \(\tilde{\alpha}\) with a fixed \(\theta=0.03\). (b) \(\mathcal{T}_H\) for different values of \(\theta\) with a fixed \(\tilde{\alpha}=0.03\). The results show that even in the simultaneous presence of both corrections, both \(\tilde{\alpha}\) and \(\theta\) remain relevant to the thermal structure.}
\label{fig:temperature_Combinaded_effects}
\end{figure}

\FloatBarrier

{As shown in Figs. (\ref{fig:temperature_individual_effects}) and (\ref{fig:temperature_Combinaded_effects}), both Gauss-Bonnet and noncommutative corrections primarily impact the small-horizon regime. As \(r_h\) increases, the corrected temperature profiles gradually approach classical behavior. On the other hand, near the minimum allowed radius, the deformation parameters significantly alter both the configuration and the magnitude of the maximum temperature, before it drops to zero. Thus, these parameters result in the elimination of the Hawking temperature singularity.
Furthermore, from the condition \(d\mathcal{T}_H/dr_h=0\), one arrives at \(r_{min}=\sqrt{\theta} +\sqrt{\theta +3 \tilde{\alpha}/2 }\).}

{Given \(r_h=r_{\min}=\sqrt{\theta} +\sqrt{\theta +3\tilde{\alpha}/2}\), the maximum temperature associated with this configuration is expressed by}
\begin{eqnarray}
    \mathcal{T}_H = \dfrac{1}{2\pi r_{min}}\left( \dfrac{1}{2} + \dfrac{\tilde{\alpha}}{4 r_{min}^2}\right).
\end{eqnarray}
{Setting \(\theta= \tilde{\alpha} = 0.03\), we find \(r_{min} \approx 0.447066\) and a maximum temperature \(\mathcal{T}_{max}\approx0.191358\). Thermodynamic analysis can therefore be applied to entropy, which is calculated from}
\begin{eqnarray}
 S=\int \frac{1}{\mathcal{T}_{H}}\frac{\partial \Tilde{M}}{\partial{r}_h} d{r}_h,  
\end{eqnarray}
where, from Eq. (\ref{eq:rplus-compact}), we have
\begin{eqnarray}
    \frac{\partial \Tilde{M}}{\partial{r}_h}\approx \dfrac{2M_0}{r_h} \left( 1+ \dfrac{\tilde{\alpha}}{r^2_h} - \dfrac{\sqrt{\theta}}{2r_h}\right).
\end{eqnarray}
So, we obtain
\begin{eqnarray}
\mathcal{S}&=&\frac{S}{M_0}=\int\frac{2}{r_h\mathcal{T}_{H}}\left(1 + \dfrac{\tilde{\alpha}}{r^2_h} - \dfrac{\sqrt{\theta}}{2r_h} \right)d{r}_h
=4\pi\int \left[\left(1 + \dfrac{\tilde{\alpha}}{r^2_h} - \dfrac{\sqrt{\theta}}{2r_h}\right)\left(1 + \dfrac{\tilde{\alpha}}{2r^2_h} +\frac{\sqrt{\theta}}{r_h}\right)\right]dr_h,
\\
&=&4\pi r_h - \dfrac{3 \tilde{\alpha}}{r_h}+ 2\pi\sqrt{\theta}\ln r_h.
\end{eqnarray}
As a result, the entropy acquires a logarithmic correction associated with the noncommutativity parameter. In the limit \(\theta \to 0\) and \(\tilde{\alpha} \to 0\), the contributions vanish, and the standard expression for BTZ black hole entropy is recovered.
Thermal stability can be assessed through specific heat capacity, represented as
\begin{eqnarray}
C=\frac{\partial\Tilde{M}}{\partial r_h}\left(\frac{\partial\mathcal{T}_H}{\partial r_h} \right)^{-1}.    
\end{eqnarray}   
Hence, we find
\begin{eqnarray}
    \mathcal{C}=\frac{C}{M_0}=-4\pi r_h \left( 1 + \dfrac{5\tilde{\alpha}}{2r^2_h} + \dfrac{3\sqrt{\theta}}{2r_h}\right)\left(1 + \dfrac{3\tilde{\alpha}}{2 r^2}+\dfrac{2\sqrt{\theta}}{r_h}\right)\left(1 - \dfrac{3\tilde{\alpha}}{2r^2_h} -\dfrac{2\sqrt{\theta}}{r_h} +\cdots \right).
\end{eqnarray}
\begin{figure}[htbh!]
\centering
\subfigure[]{\includegraphics[scale=0.22]{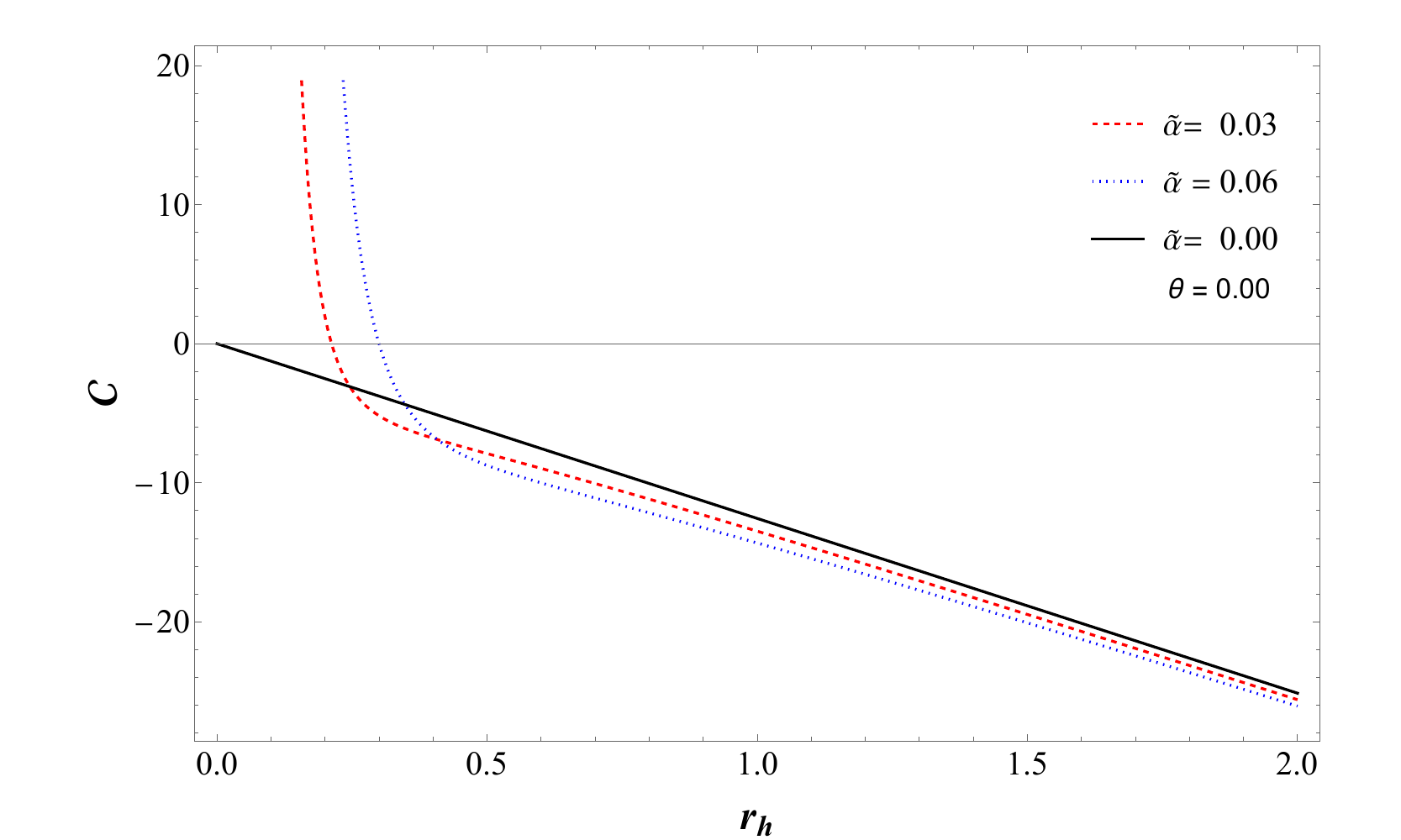}}
\subfigure[]{\includegraphics[scale=0.22]{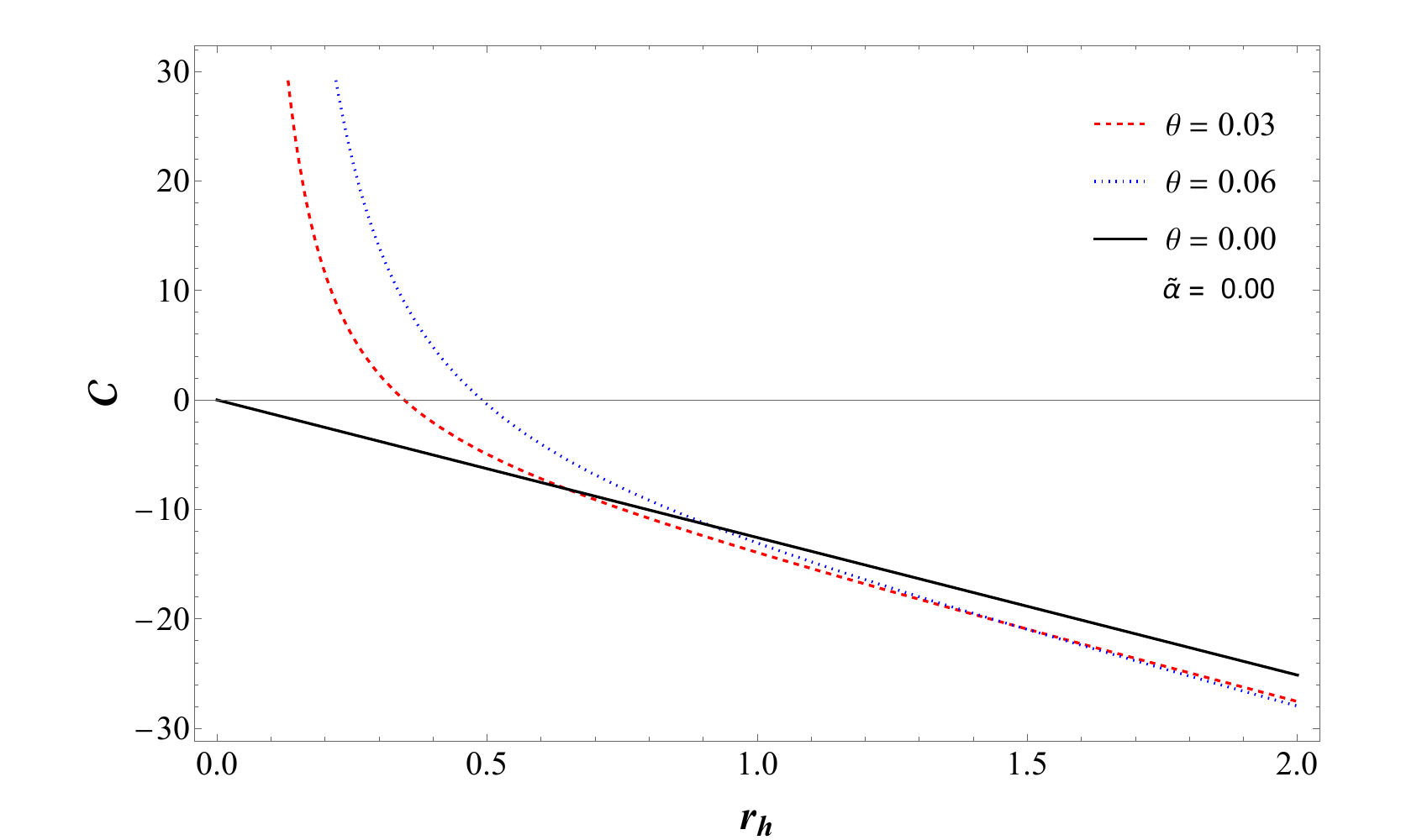}}
 \caption{Specific heat capacity $\mathcal{C}$ as a function of $r_h$ in the presence of Gauss-Bonnet and noncommutative corrections, considered separately. (a) Different values of $\tilde{\alpha}$ for $\theta=0$; (b) different values of $\theta$ for $\tilde{\alpha}=0$. In both scenarios, the corrections establish regions with $\mathcal{C}>0$ and $\mathcal{C}<0$, modifying the thermal stability structure in the small-horizon regime and indicating the presence of a remnant.}
\label{fig:heat_capacity_individual}
\end{figure}
\begin{figure}[htbh!]
\centering
\subfigure[]{\includegraphics[scale=0.22]{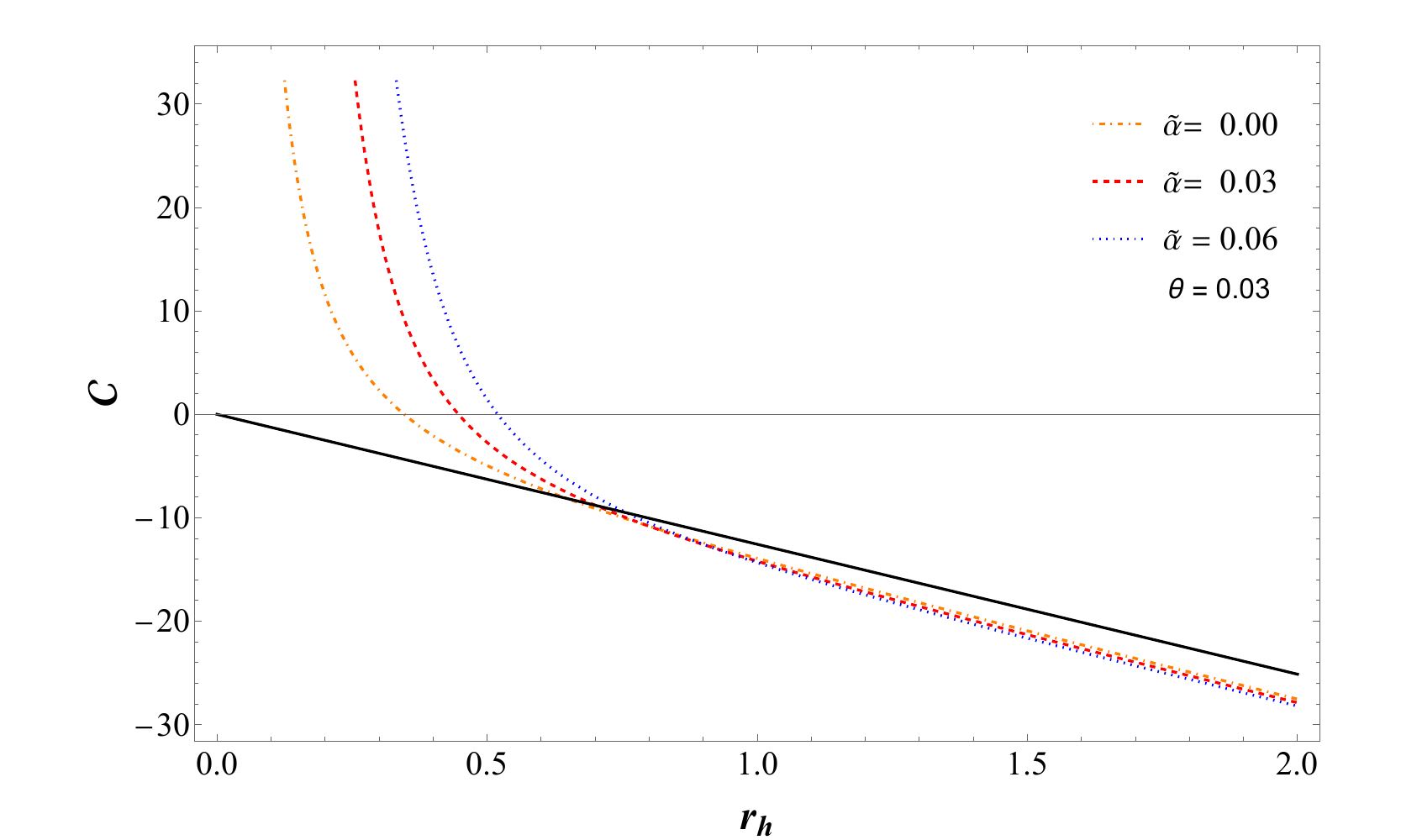}}
\subfigure[]{\includegraphics[scale=0.22]{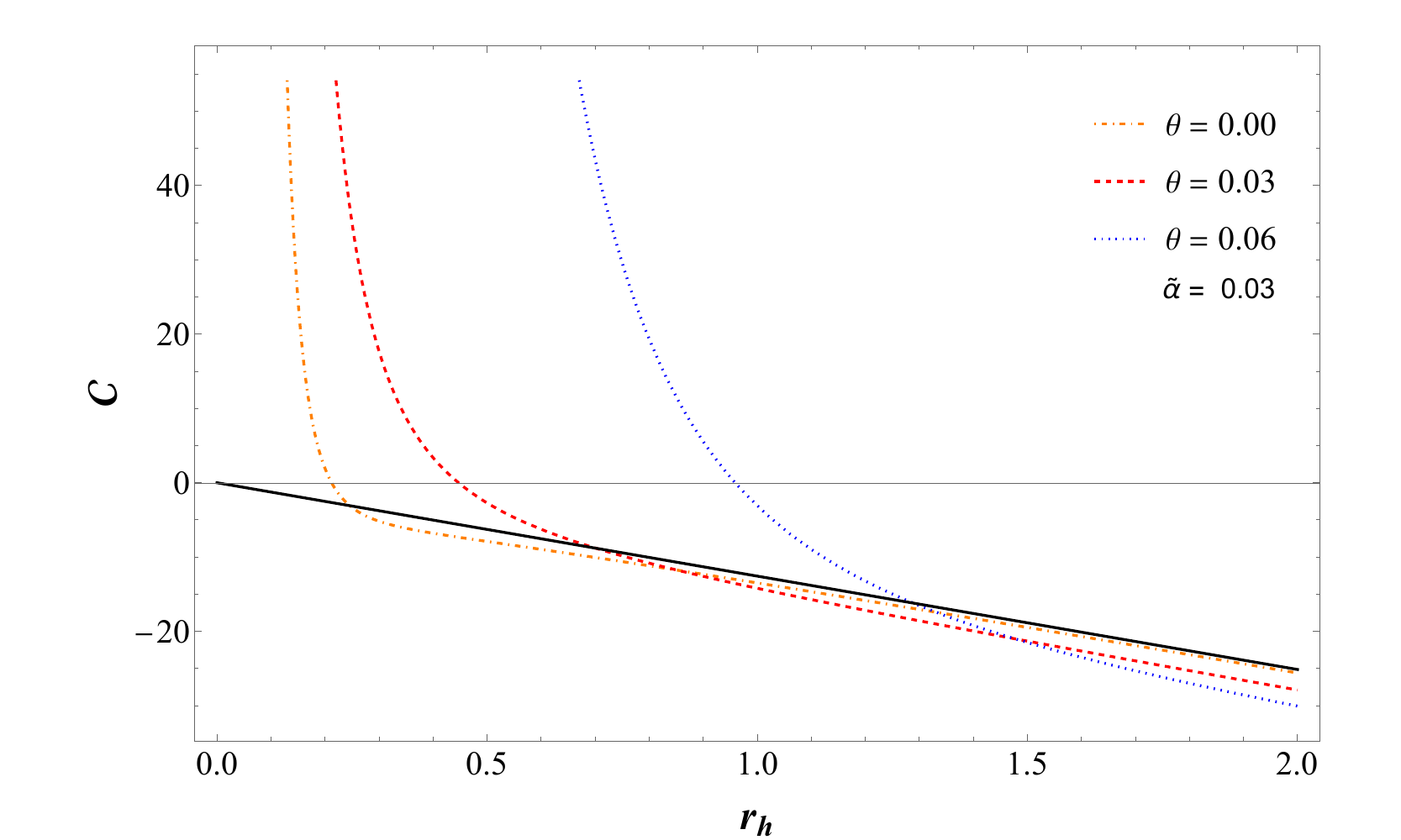}}
 \caption{Specific heat capacity $\mathcal{C}$ as a function of the horizon radius $r_h$ in the presence of Gauss-Bonnet and noncommutative corrections. (a) Specific heat capacity $\mathcal{C}$ for different values of \(\tilde{\alpha}\), with \(\theta=0.03\). (b) Specific heat capacity $\mathcal{C}$ for different values of \(\theta\), with \(\tilde{\alpha}=0.03\). The corrections introduce a region with $\mathcal{C}>0$ in the small-horizon sector, associated with locally stable configurations and consistent with the remnant structure indicated by the Hawking temperature behavior.}
\label{fig:heat_capacity_combined}
\end{figure}

Figs.(\ref{fig:heat_capacity_individual}) and
(\ref{fig:heat_capacity_combined}) indicate that the addition of Gauss-Bonnet and noncommutative corrections significantly alters the thermal stability in the small-horizon sector. Specifically, the emergence of a branch with $\mathcal{C}>0$ characterizes a locally stable region. Given that the Hawking temperature drops to zero at a finite radius, the results provide thermodynamic support for the existence of a remnant configuration, without the need to extend the evolution down to $r_h=0$.

For $r_h = r_{min}=\sqrt{\theta} +\sqrt{\theta +3 \tilde{\alpha}/2 }$, the thermal evolution reaches its limiting configuration, associated with the formation of a black hole remnant. Furthermore, using $r_{min}=\sqrt{l^2M_{min}}$, the corresponding minimum mass is given by
\begin{eqnarray}
M_{min}
&=&
\frac{r_{min}^{2}}{l^{2}} = \frac{3\tilde{\alpha} + 8\theta}{2l^{2}} = -\Lambda\left(\frac{3}{2}\tilde{\alpha} + 4\theta\right).
\end{eqnarray}
Thus, for $M_{min}>0$ and considering $\tilde{\alpha}>0$, a negative cosmological constant, $\Lambda<0$, is required. Finally, in the limit $\theta=\tilde{\alpha}=0$, we recover $\mathcal{C}=-4\pi r_h$, corresponding to the result without the corrections considered.
\section{STRUCTURE EQUATIONS OF NONCOMMUTATIVE GRAVASTARS}\label{sec2}

To construct the gravastar configuration, we considered two different types of spacetime manifolds. The interior region is represented by the manifold $M_{-}$, and the exterior geometry is described by $M_{+}$. A thin layer situated on a hypersurface $\Sigma$ separates these two regions, where matching conditions are applied. Through the cut-and-paste formalism, the entirety of spacetime is obtained by joining the interior and exterior manifolds via the junction surface $\Sigma$. 
The exterior region is described by the Gauss-Bonnet-modified BTZ geometry in \((2+1)\) dimensions~\cite{Hennigar:2020fkv}, with line element
\begin{eqnarray}
    ds^{2} = -f(r)_{+} dt_{+}^{2} + f(r)_{+}^{-1}dr_{+}^{2} + r_{+}^{2}d\phi_{+}^{2},
\end{eqnarray}
where $t$ is the physical time in the outer region, and
\begin{eqnarray}
\label{frl}
f(r)_+= \dfrac{r^{2}}{l^{2}} - M_0 - \dfrac{\alpha}{r^{2}}\left( \dfrac{r^{2}}{l^{2}}-M_0\right)^{2} = M -\Lambda r^{2} - \dfrac{\alpha}{r^{2}}\left( M - \Lambda r^{2} \right)^{2}.
\end{eqnarray}
Here $M=-M_0$ and $\Lambda=-1/l^2$ is the cosmological constant.

We can rewrite the metric function (\ref{frl}) in Schwarzschild-type form~\cite{Silva:2023bsu, Anacleto:2025oef}:
\begin{eqnarray}
f(r)_{+} = 1- \frac{b_{+}}{r},
\end{eqnarray}
being
\begin{eqnarray}
b_+=-r\left[M -\Lambda r^{2} - \dfrac{\alpha}{r^{2}}\left( M - \Lambda r^{2} \right)^{2}\right] +r.
\end{eqnarray}
For the inner region, the line element describing the three-dimensional spacetime Gauss-Bonnet with noncommutative geometry is given by  
\begin{eqnarray}
ds^{2} =  - g(r)_{-}dt_{-}^{2} + f(r)_{-}^{-1}dr_{-}^{2} + r_{-}^{2}d\phi_{-}^{2},
\label{btzcm}
\end{eqnarray}
where the metric functions $g(r)_{-}$ and $f(r)_{-}$ are given respectively by
\begin{eqnarray}
g(r)_{-}&=& 
M\left(1 - \frac{\sqrt{\theta}}{\sqrt{r^2 + \theta}} \right)\left(1 + 2\alpha \Lambda\right) - \Lambda r^{2}\ \left(1 + \alpha \Lambda\right) - \frac{\alpha M^2}{r^2}\left(1 - \frac{2\sqrt{\theta}}{\sqrt{r^2 + \theta}}\right) .
\label{a05}
\\
f(r)_{-}&=& 1- \frac{b_{-}}{r}, 
\end{eqnarray}      
being
\begin{eqnarray}
b_{-} = -r   \left[M\left( 1- \frac{\sqrt{\theta}}{\sqrt{r^{2}+\theta}}\right)-\frac{\alpha M^{2}}{r^{2}}\left( 1- \frac{2\sqrt{\theta}}{\sqrt{r^{2}+\theta}}\right) \right]  + r,
\end{eqnarray}
and $\pm$ represents the outer and inner geometry, respectively. 

Our objective is to match the manifolds $M_{-}$ and $M_{+}$ at their common boundary, thereby constructing a single spacetime manifold $M$, such that  $M=M_{-}\cup M_{+}$, with $\Sigma=\Sigma_{-}=\Sigma_{+}$~\cite{Lobo:2015lbc}. We use the intrinsic metric induced on the junction hypersurface 
to calculate the surface energy density and pressure related to the thin shell, respectively.
\begin{eqnarray}
ds_{\Sigma}^{2}=-d\tau^{2}+a(\tau)^{2}
d\phi^{2},
\end{eqnarray}
where $\tau$ denotes the proper time measured by an observer comoving with the shell in the $(2+1)$-dimensional spacetime considered here.

The junction surface is described by $x^{\nu} (\tau,\phi) = (t(\tau), a(\tau), \phi)$, where the unit normal vectors associated with this surface are given by ~\cite{MartinMoruno:2011rm}:
\begin{eqnarray}
    n^{\mu}_{+} = \left( \dfrac{1}{M -\Lambda a^2 - \frac{\alpha}{a^2} (M-\Lambda a^{2})^{2}}\dot{a}, \sqrt{M -\Lambda a^2 - \frac{\alpha}{a^2} (M-\Lambda a^{2})^{2} + \dot{a}^{2}},0\right),
\end{eqnarray}
and
\begin{eqnarray}
    n^{\mu}_{-} = \left(\frac{1}{{M}_{\theta}-\dfrac{\alpha M^{2}_{\theta}}{a^{2}}}\dot{a}, \sqrt{M_{\theta}-\frac{\alpha M^{2}_{\theta}}{a^{2}} +\dot{a}^{2}},0
    \right),
\end{eqnarray}
where an overdot denotes differentiation with respect to $\tau$. The normal vectors satisfy the condition $n^{\mu}n_{\mu} = 1$ for the circularly symmetric \((2+1)\)-dimensional spacetime considered here. The following expression is employed to calculate extrinsic curvatures ~\cite{Visser:1995cc, MartinMoruno:2011rm}.
\begin{eqnarray}
K_{ij}^{\pm}=-n_{\mu}\left(\frac{\partial^{2}x^{\mu}}{\partial\xi^{i}\,\partial\xi^{j}}+\Gamma_{\;\;\alpha\beta}^{\mu\pm}\;\frac{\partial x^{\alpha}}{\partial\xi^{i}}\,\frac{\partial x^{\beta}}{\partial\xi^{j}}\right)\,.\label{extrinsiccurv}
\end{eqnarray}
So the extrinsic curvatures are found as follows: 
\begin{eqnarray}
K^{\phi +}_{\,\ \phi} = \frac{1}{a}\sqrt{M -\Lambda a^2 - \frac{\alpha}{a^2} (M-\Lambda a^{2})^{2} + \dot{a}^{2}},
\end{eqnarray}
\begin{eqnarray}
    K^{\tau +}_{\,\ \tau} = \left\lbrace \frac{\ddot{a} - \Lambda {a}^2 +\dfrac{M^{2}\alpha}{a^{3}}-\Lambda^{2}a^{2} \alpha}{\sqrt{M -\Lambda a^2 - \frac{\alpha}{a^2} (M-\Lambda a^{2})^{2}  + \dot{a}^{2}}}\right\rbrace, 
\label{exouttau}
\end{eqnarray}
\begin{eqnarray}
K_{\;\;\phi}^{\phi\;-} & = & \frac{1}{a}\,\sqrt{M\left( 1- \frac{\sqrt{\theta}}{\sqrt{a^{2}+\theta}}\right)-\frac{\alpha M^{2}}{a^{2}}\left( 1- \frac{2\sqrt{\theta}}{\sqrt{a^{2}+\theta}}\right) +\dot{a}^{2}}\;,\label{genKplustheta}
\end{eqnarray}
\begin{eqnarray}
K_{\;\;\tau}^{\tau\;-} & = & \left\{ \frac{ \ddot{a} - \dfrac{M\left[a^4 \sqrt{\theta} + 2Ma^{2} \alpha\left( \sqrt{a^2 +\theta} -3\sqrt{\theta}\right) +2M\alpha\theta \left(\sqrt{a^2+\theta}-2\sqrt{\theta} \right) \right]}{2a^3(a^{2} +\theta)^{3/2}}}{ \sqrt{ M\left( 1- \frac{\sqrt{\theta}}{\sqrt{a^{2}+\theta}}\right)-\frac{\alpha M^{2}}{a^{2}}\left( 1- \frac{2\sqrt{\theta}}{\sqrt{a^{2}+\theta}}\right) +\dot{a}^{2}} }\right\} \ .
\label{genKminustautau}
\end{eqnarray}
The discontinuity in the extrinsic curvature across the junction hypersurface $\Sigma$ is defined by $k^{i}_{\, j}=K^{i\, +}_{\, j} - K^{i\, -}_{\, j}$. According to the Lanczos-Israel junction conditions, the surface energy-momentum tensor on $\Sigma$ is given by~\cite{Lobo:2015lbc}
\begin{eqnarray}
S^{i}_{\, j} = - \frac{1}{8 \pi} \left(k^{i}_{\, j} - \delta^{i}_{\, j} \,\ k^{k}_{\,\, k}\right).
\end{eqnarray}
Assuming the matter distribution on the hypersurface is described by a perfect fluid confined to the shell, the surface energy-momentum tensor in (2+1)-dimensional spacetime can be written as $S^{i}_{\,j} = diag( - \sigma, \mathcal{P})$ where 
$\sigma=- {K^{\phi }_{\,\ \phi}}/{4 \pi }$ is the surface density 
and $\mathcal{P}={(K^{\tau }_{\,\ \tau} + K^{\phi }_{\,\ \phi} )}/{8 \pi }$ is the surface pressure~\cite{Perry:1991qq}. These are given as follows:

\begin{eqnarray}
    \sigma = - \dfrac{1}{4 \pi a} 
\left[ \sqrt{M -\Lambda a^2 - \frac{\alpha}{a^2} (M-\Lambda a^{2})^{2} + \dot{a}^{2}} - \sqrt{M_{\theta}
-\frac{\alpha M^{2}_{\theta}}{a^{2}} +\dot{a}^{2}} \, \right], 
\end{eqnarray}
\begin{eqnarray}
\mathcal{P} &=&
\frac{1}{8 \pi a} \left[\dfrac{\dot{a}^{2} + \ddot{a}  + M (1- 2\alpha\Lambda) -  2 \Lambda a^{2}(1 -\alpha\Lambda) }{\sqrt{M -\Lambda a^2 - \dfrac{\alpha}{a^2} (M-\Lambda a^{2})^{2}  + \dot{a}^{2}}} -\frac{\dot{a}^{2}+ \ddot{a} + M\left( 1 -\dfrac{\sqrt{\theta}(a^{2} +2 \theta)}{2 (a^{2} + \theta)^{3/2}}\right)- \dfrac{M^{2}\alpha \sqrt{\theta}}{(a^{2} +\theta)^{3/2}}}{ \sqrt{ M_{\theta}-\dfrac{\alpha M^{2}_{\theta}}{a^{2}}
 +\dot{a}^{2}} } \right].
\end{eqnarray}
From Eqs. (\ref{exouttau}) and (\ref{genKminustautau}), one obtains:
\begin{eqnarray}
\sigma+2\mathcal{P} =\dfrac{\kappa_{\;\tau}^{\tau}}{4\pi}
=\frac{1}{4\pi~a}\left[ 
\frac{\ddot{a} - \Lambda {a}^2 (1+\alpha\Lambda) +\dfrac{M^{2}\alpha}{a^{2}} }{\sqrt{M -\Lambda a^2 - \frac{\alpha}{a^2} (M-\Lambda a^{2})^{2}  + \dot{a}^{2}}} -  \frac{ \ddot{a} -  M \left(\dfrac{2Ma^{2} \alpha \sqrt{a^2 +\theta} }{2a^2(a^{2} +\theta)^{3/2}}+ P_{\theta}\right)}{ \sqrt{ M_{\theta}-\dfrac{\alpha M^{2}_{\theta}}{a^{2}}
 +\dot{a}^{2}} } \right],
\label{s2P}
\end{eqnarray}
where
\begin{eqnarray}
    P_{\theta} = \dfrac{\left[a^4 \sqrt{\theta} - 6Ma^{2} \alpha\sqrt{\theta} +2M\alpha\theta \left(\sqrt{a^2+\theta}-2\sqrt{\theta} \right) \right]}{2a^2(a^{2} +\theta)^{3/2}}.
\end{eqnarray}
To investigate the static equilibrium configurations of the gravastar, we restrict our analysis to static configurations satisfying $a_{0} \in (r_{-},r_{+})$. Under this assumption, we can be written as:
\begin{eqnarray}
    \sigma(a_0) = - \dfrac{1}{4 \pi a_0} 
\left[ \sqrt{M -\Lambda a_0^2 - \frac{\alpha}{a_0^2} (M-\Lambda a_{0}^{2})^{2}} - \sqrt{M\left( 1- \dfrac{\sqrt{\theta}}{\sqrt{a_0^{2}+\theta}}\right)-\dfrac{\alpha M^{2}}{a_0^{2}}\left( 1- \dfrac{2\sqrt{\theta}}{\sqrt{a_0^{2}+\theta}}\right) } \, \right], 
\end{eqnarray}
\begin{eqnarray}
\mathcal{P}(a_0) &=&
\frac{1}{8 \pi a_{0}} \left[\dfrac{ M -  2 \Lambda a_{0}^{2} + 2a_{0}^{2}\alpha\Lambda^{2} - 2M \alpha \Lambda}{\sqrt{M -\Lambda a_{0}^2 - \dfrac{\alpha}{a_{0}^2} (M-\Lambda a_{0}^{2})^{2} }} -\frac{ M\left( 1 -\dfrac{\sqrt{\theta}(a_{0}^{2} +2 \theta)}{2 (a_{0}^{2} + \theta)^{3/2}}\right) - \dfrac{M^{2}\alpha \sqrt{\theta}}{(a_{0}^{2} +\theta)^{3/2}}}{ \sqrt{ M\left( 1 - \dfrac{\sqrt{\theta}}{\sqrt{a_{0}^{2}+\theta}}\right)-\dfrac{\alpha M^{2}}{a_{0}^{2}}\left( 1- \dfrac{2\sqrt{\theta}}{\sqrt{a_{0}^{2}+\theta}}\right)} } \right],
\end{eqnarray}
\begin{eqnarray}
\sigma(a_0)+2\mathcal{P}(a_0) = \frac{1}{4\pi~a_{0}}\left[ 
\frac{\dfrac{M^{2}\alpha}{a^{2}_{0}}- \Lambda {a}_{0}^2 -\Lambda^{2}a_{0}^{2} \alpha}{\sqrt{M -\Lambda a^2_{0} - \dfrac{\alpha}{a^2_{0}} (M-\Lambda a_{0}^{2})^{2}}}  +\frac{ M \left(\dfrac{2Ma_{0}^{2} \alpha \sqrt{a_{0}^2 +\theta} }{2a^2(a_{0}^{2} +\theta)^{3/2}} + P_{\theta}\right)}{ \sqrt{ M\left( 1- \dfrac{\sqrt{\theta}}{\sqrt{a_{0}^{2}+\theta}}\right)-\dfrac{\alpha M^{2}}{a^{2}_{0}}\left( 1- \dfrac{2\sqrt{\theta}}{\sqrt{a^{2}_{0}+\theta}}\right)} } \right].
\label{s2Padm}
\end{eqnarray}
Now, we can express the above equations in terms of the dimensionless parameter $\Tilde{\Lambda} = \Lambda a^2_0$, $\gamma={\alpha}/{a_{0}^{2}}$ and $\Theta ={\sqrt{\theta}}/{a_{0}}$
\begin{eqnarray}
    \Tilde{\sigma} = - \dfrac{1}{4 \pi } 
\left[ \sqrt{M -\Tilde{\Lambda} - \gamma (M-\Tilde{\Lambda})^{2}} - \sqrt{M\left( 1- \frac{\Theta}{\sqrt{1+\Theta^{2}}}\right)-\gamma M^{2}\left( 1- \frac{2\Theta}{\sqrt{1+\Theta^{2}}}\right) } \, \right],
\label{sigadm}
\end{eqnarray}
\begin{eqnarray}
\tilde{\mathcal{P}} &=& \frac{1}{8 \pi } \left[\frac{M -2 \Tilde{\Lambda} + 2\gamma\Tilde{\Lambda}^{2} - 2M\gamma \tilde{\Lambda}}{\sqrt{M -\Tilde{\Lambda} - \gamma (M-\Tilde{\Lambda})^{2}}} -\dfrac{ M \left(1 - \dfrac{\Theta (1+ 2\Theta^{2})}{2(1+ \Theta^{2})^{3/2}}\right) - \dfrac{M^{2}\gamma \Theta}{(1 + \Theta^{2})^{3/2}} }{ \sqrt{ M\left( 1- \dfrac{\Theta}{\sqrt{1+\Theta^{2}}}\right)-\gamma M^{2}\left( 1- \dfrac{2\Theta}{\sqrt{1 +\Theta^{2}}}\right)} } \right],
\label{presadm}
\end{eqnarray}
\begin{eqnarray}
\tilde{\sigma}+2\tilde{\mathcal{P}} = \frac{1}{4\pi}\left[ 
\frac{M^{2}\gamma- \tilde{\Lambda} -\tilde{\Lambda}^{2} \gamma}{\sqrt{M -\tilde{\Lambda} - \gamma (M-\tilde{\Lambda})^{2}}}  + \frac{ M \left(\dfrac{2M\gamma \sqrt{1 +\Theta^{2}} }{2(1 +\Theta^{2})^{3/2}}+ P_{\Theta}\right)}{ \sqrt{ M\left( 1- \dfrac{\Theta}{\sqrt{1+\Theta^{2}}}\right)-\gamma M^{2}\left( 1- \dfrac{2\Theta}{\sqrt{1+\Theta^{2}}}\right)} } \right],
\label{s2Ptild}
\end{eqnarray}
where the quantities $\tilde{\sigma}= a_0\sigma(a_0) $ $\Tilde{\mathcal{P}}= a_0\mathcal{P}(a_0)$, $\left[\sigma(a_{0})+2\mathcal{P}(a_{0})\right] a_0 = \Tilde{\sigma}+2\Tilde{\mathcal{P}}$ and
\begin{eqnarray}
    P_{\Theta} = \dfrac{\left[\Theta - 6M \gamma \Theta + 2M\gamma\Theta^{2} \left(\sqrt{1+\Theta^{2}}-2\Theta \right) \right]}{2\left(1 +\Theta^{2}\right)^{3/2}},
\end{eqnarray}
have been introduced.
\begin{figure}[htbh!]
\centering
\subfigure[]{\includegraphics[scale=0.17]{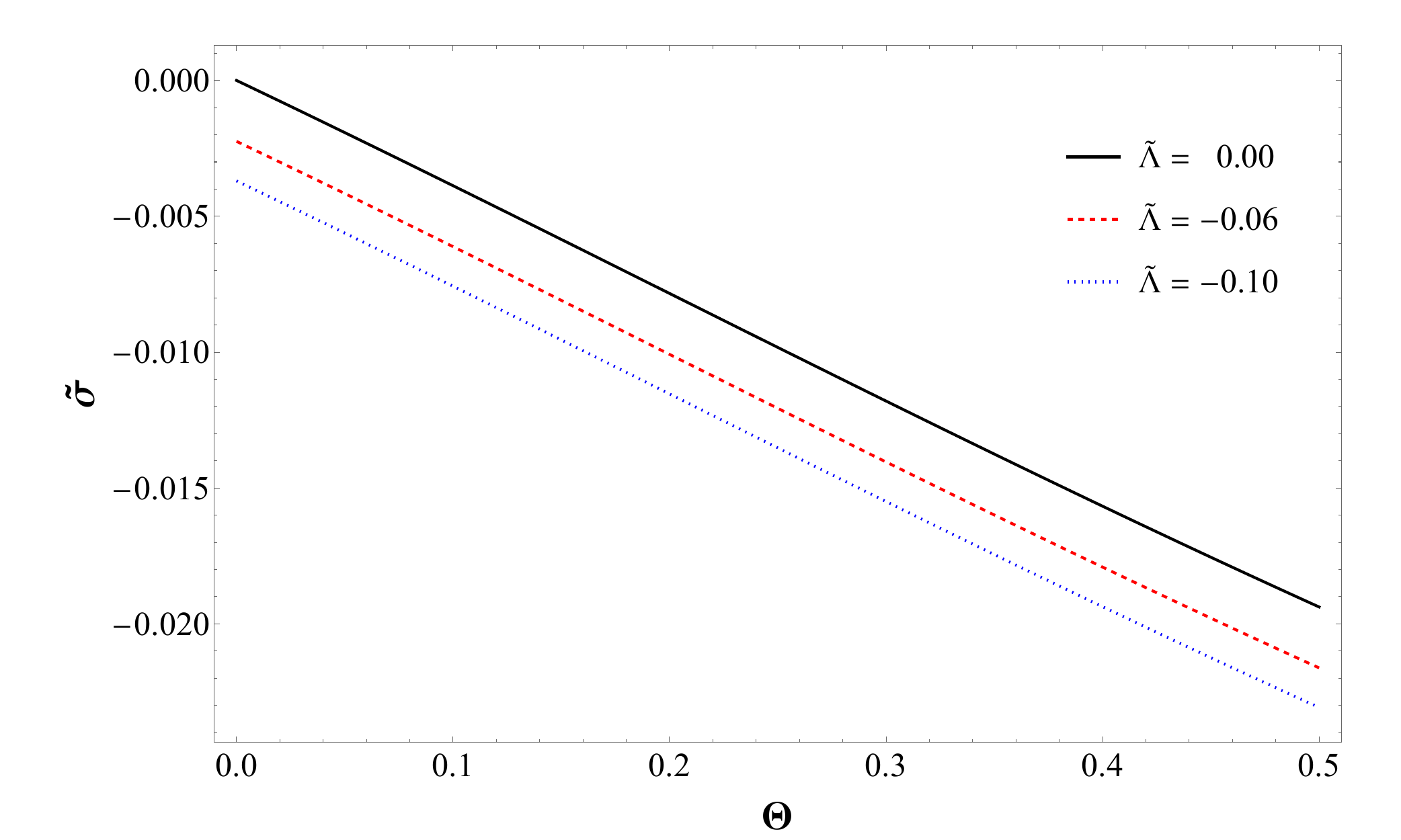}}
\subfigure[]{\includegraphics[scale=0.17]{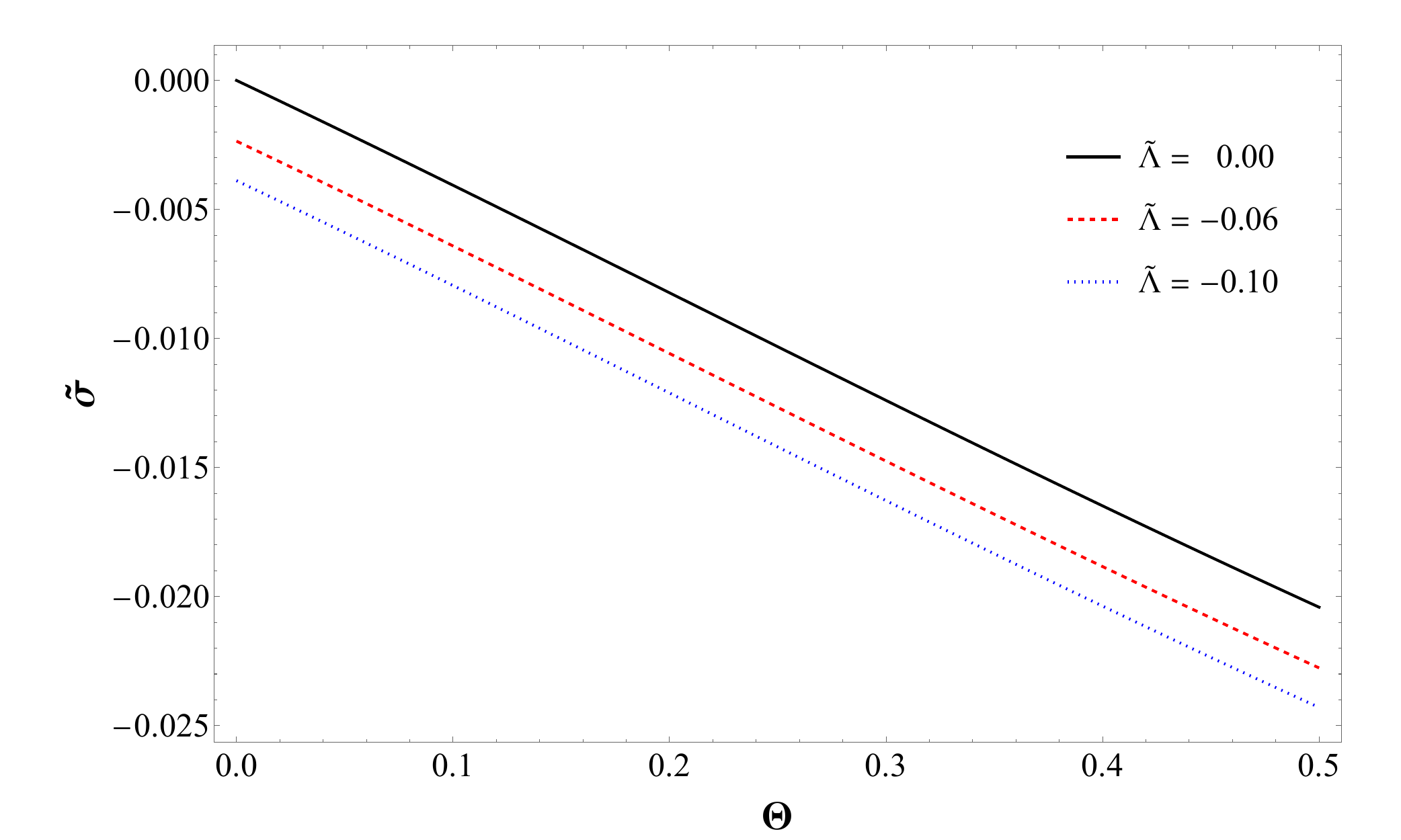}}
\caption{Energy density $\Tilde{\sigma}$ as a function of parameter $\Theta$ for $\Tilde{\Lambda} = 0$ (black, solid), $\Tilde{\Lambda} = -0.06$ (red, dashed) and $\Tilde{\Lambda} = -0.10$ (blue, dotted), assuming $ M=1 $. (a) are considered the dimensionless Gauss-Bonnet coupling $\gamma = 0.03$. (b) are considered the dimensionless Gauss-Bonnet coupling $\gamma = 0$.}
\label{Energy1}
\end{figure}
\begin{figure}[htbh!]
\centering
\subfigure[]{\includegraphics[scale=0.17]{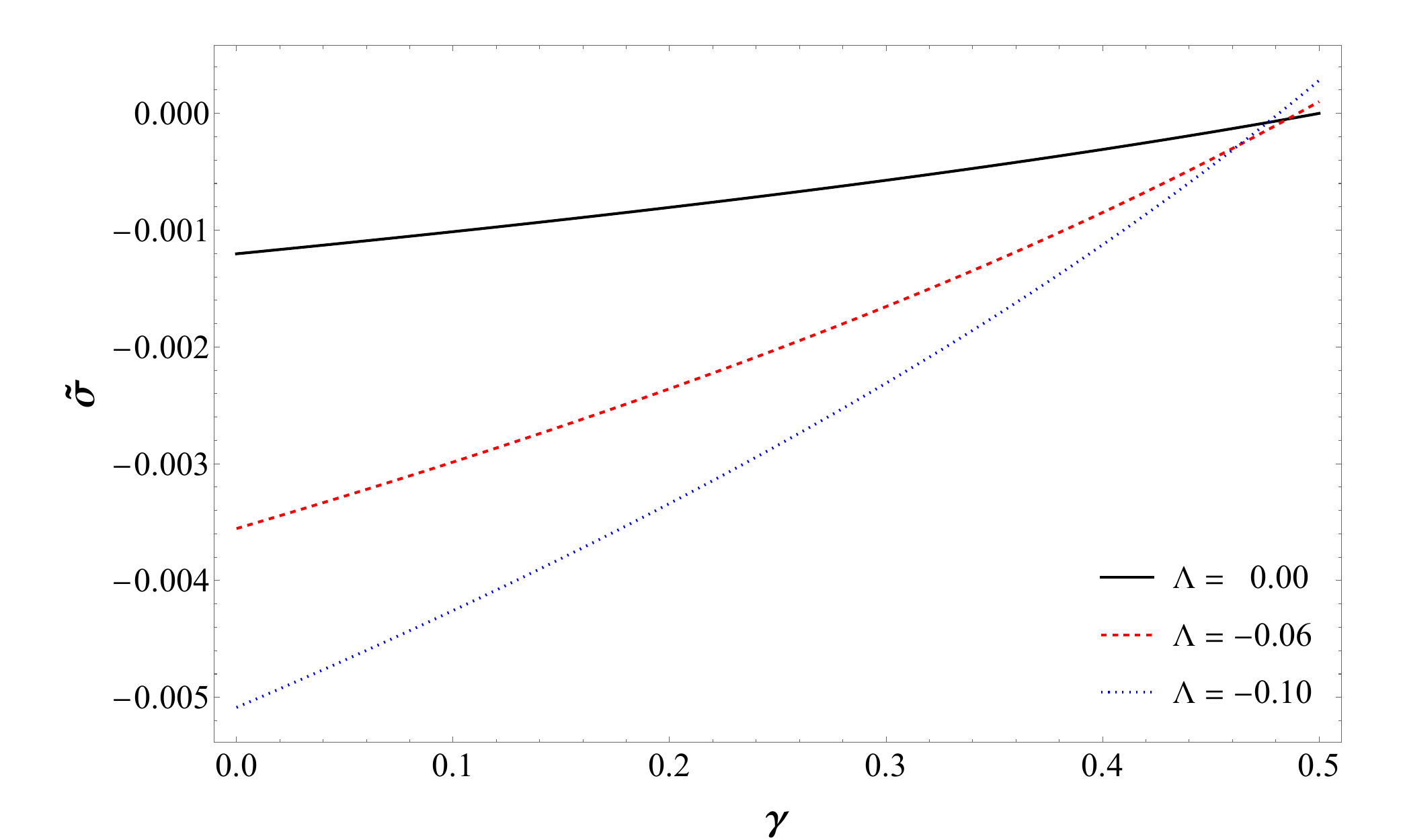}}
\subfigure[]{\includegraphics[scale=0.17]{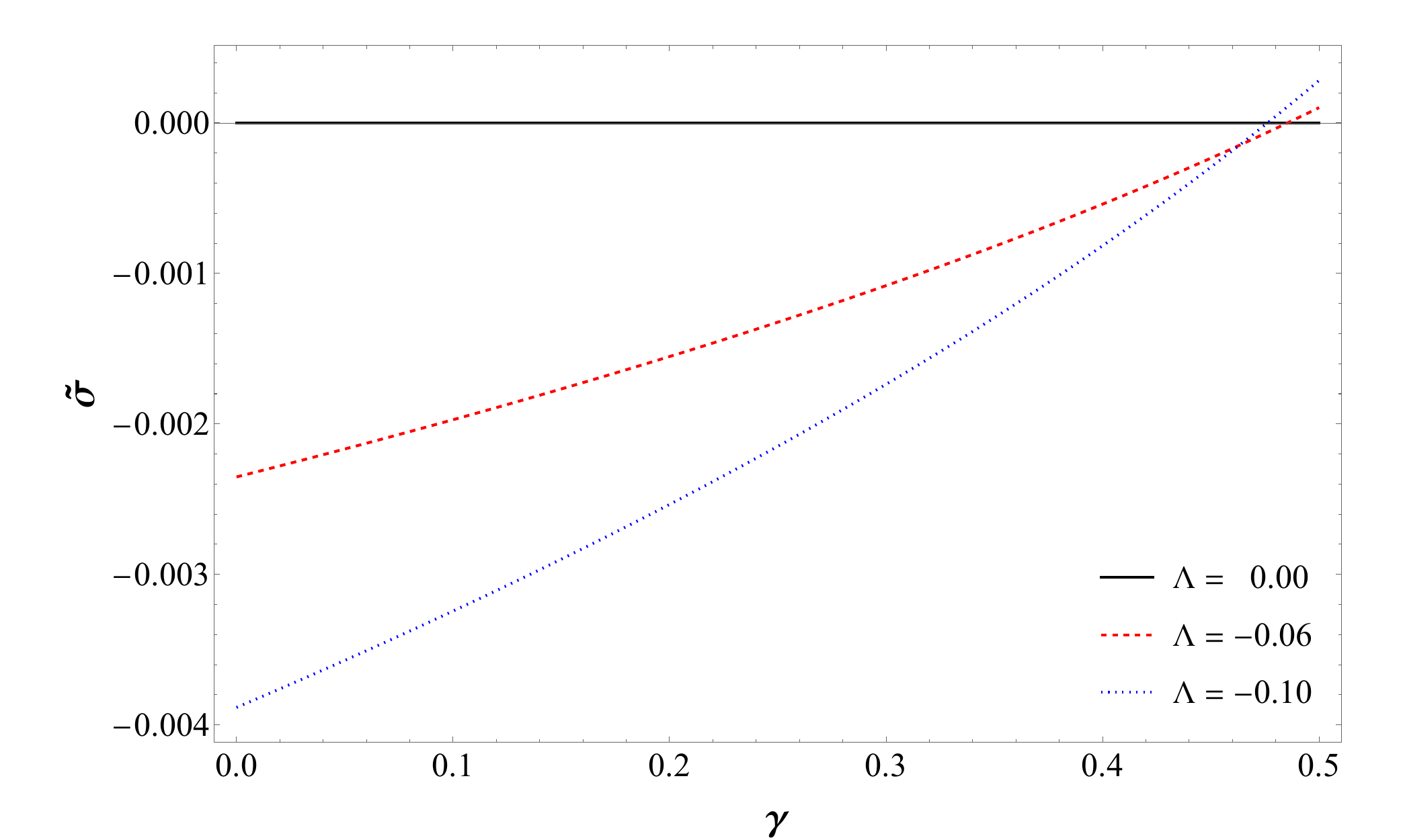}}
\caption{Energy density $\Tilde{\sigma}$ as a Gauss-Bonnet coupling $\gamma$ for $\Tilde{\Lambda} = 0$ (black, solid), $\Tilde{\Lambda} = -0.06$ (red, dashed) and $\Tilde{\Lambda} = -0.10$ (blue, dotted), assuming $ M=1 $. (a) are considered the noncommutativity parameter $\Theta=0.03$. (b) are considered the noncommutativity parameter $\Theta=0$.}
\label{Energy2}
\end{figure}
\begin{figure}[htbh!]
\centering
\subfigure[]{\includegraphics[scale=0.17]{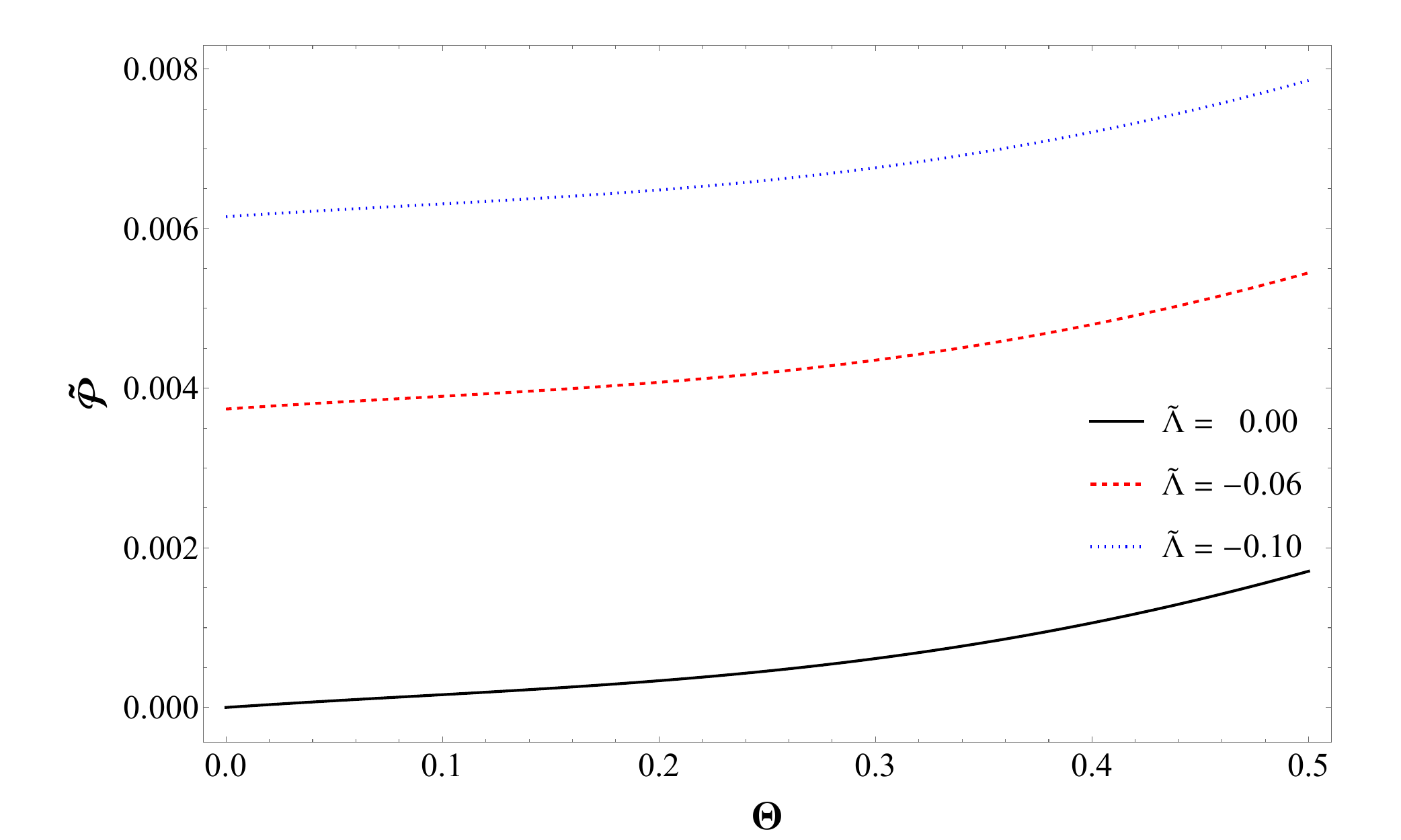}}
\subfigure[]{\includegraphics[scale=0.17]{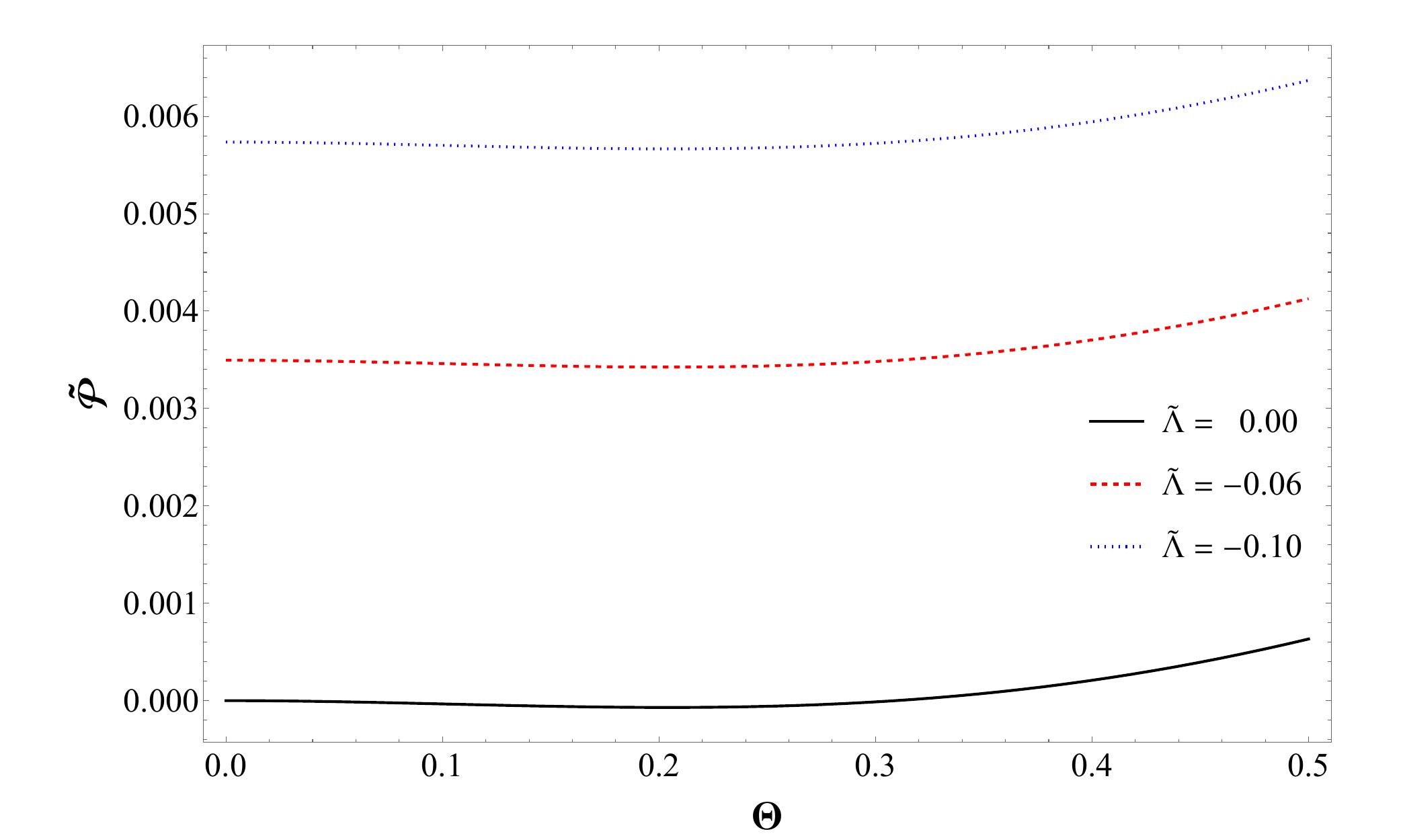}}
\caption{Pressure $\tilde{\mathcal{P}}$ as a function of parameter $\Theta$ for $\Tilde{\Lambda} = 0$ (black, solid), $\Tilde{\Lambda} = -0.06$ (red, dashed), and $\Tilde{\Lambda} = -0.10$ (blue, dotted), under the assumption that $M=1$. (a) are considered as the dimensionless Gauss-Bonnet coupling $\gamma = 0.03$. (b) are considered as the dimensionless Gauss-Bonnet coupling $\gamma = 0$.}
\label{Pres1}\end{figure}\begin{figure}[htbh!]
\centering
\subfigure[]{\includegraphics[scale=0.17]{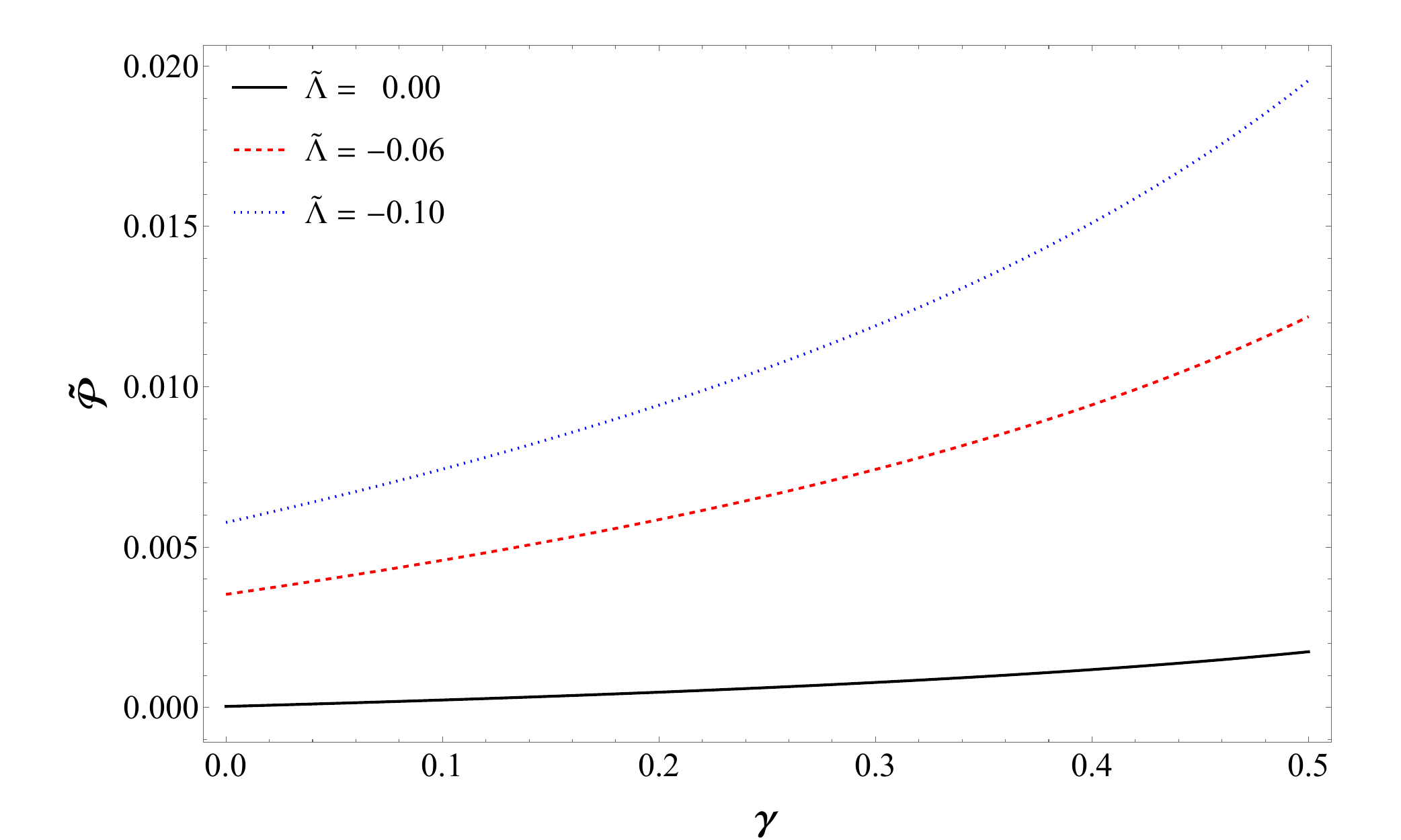}}
\subfigure[]{\includegraphics[scale=0.17]{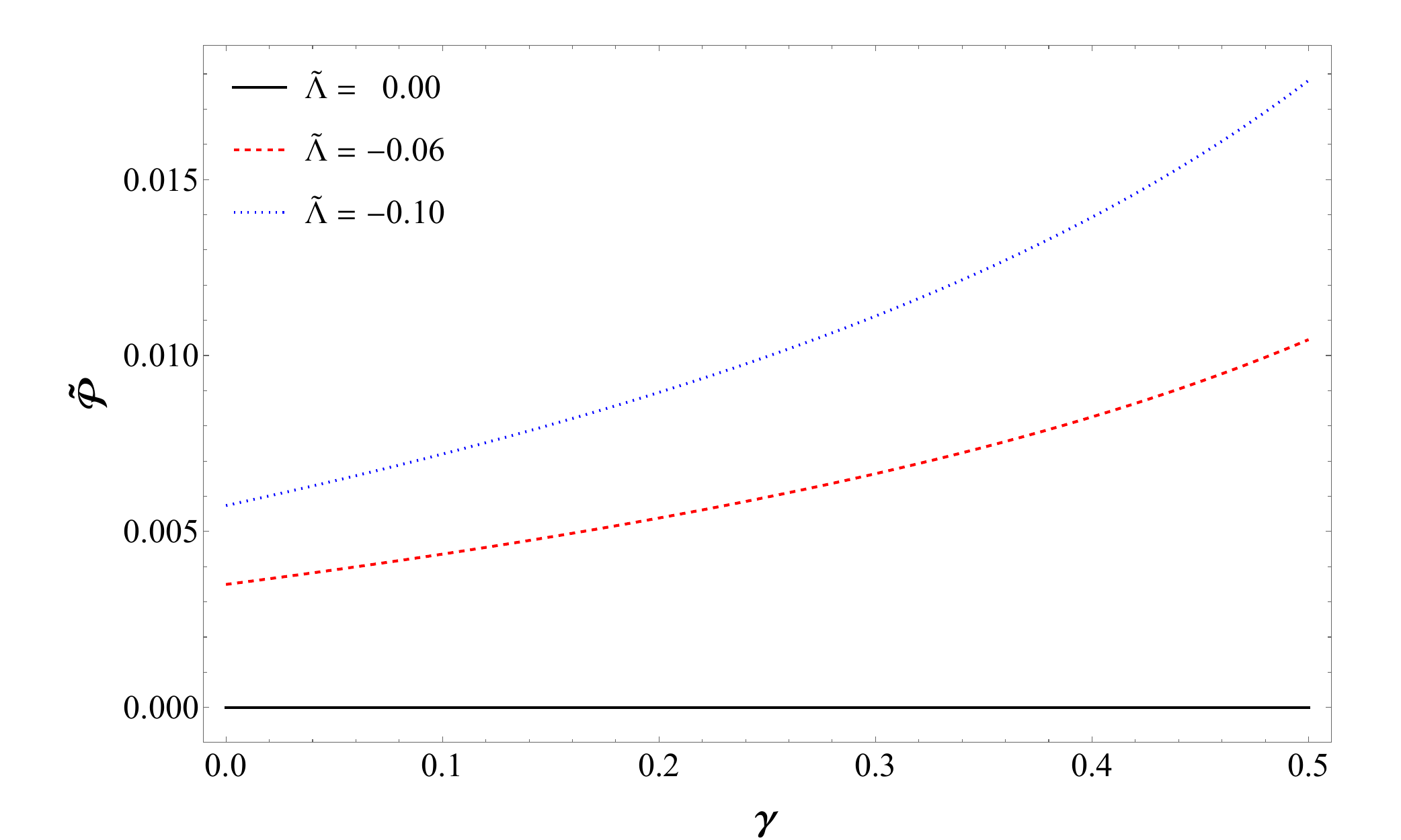}}
\caption{Pressure $\tilde{\mathcal{P}}$ as a function of the Gauss-Bonnet coupling \(\gamma\) for $\tilde{\Lambda} = 0$ (black, solid), $\Tilde{\Lambda} = -0.06$ (red, dashed), and $\Tilde{\Lambda} = -0.10$ (blue, dotted), under the assumption that $M=1$. (a) are considered the noncommutativity parameter $\Theta=0.03$. (b) are considered the noncommutativity parameter $\Theta=0$.}
\label{Pres2}
\end{figure}
\begin{figure}[htbh!]
\centering
\subfigure[]{\includegraphics[scale=0.17]{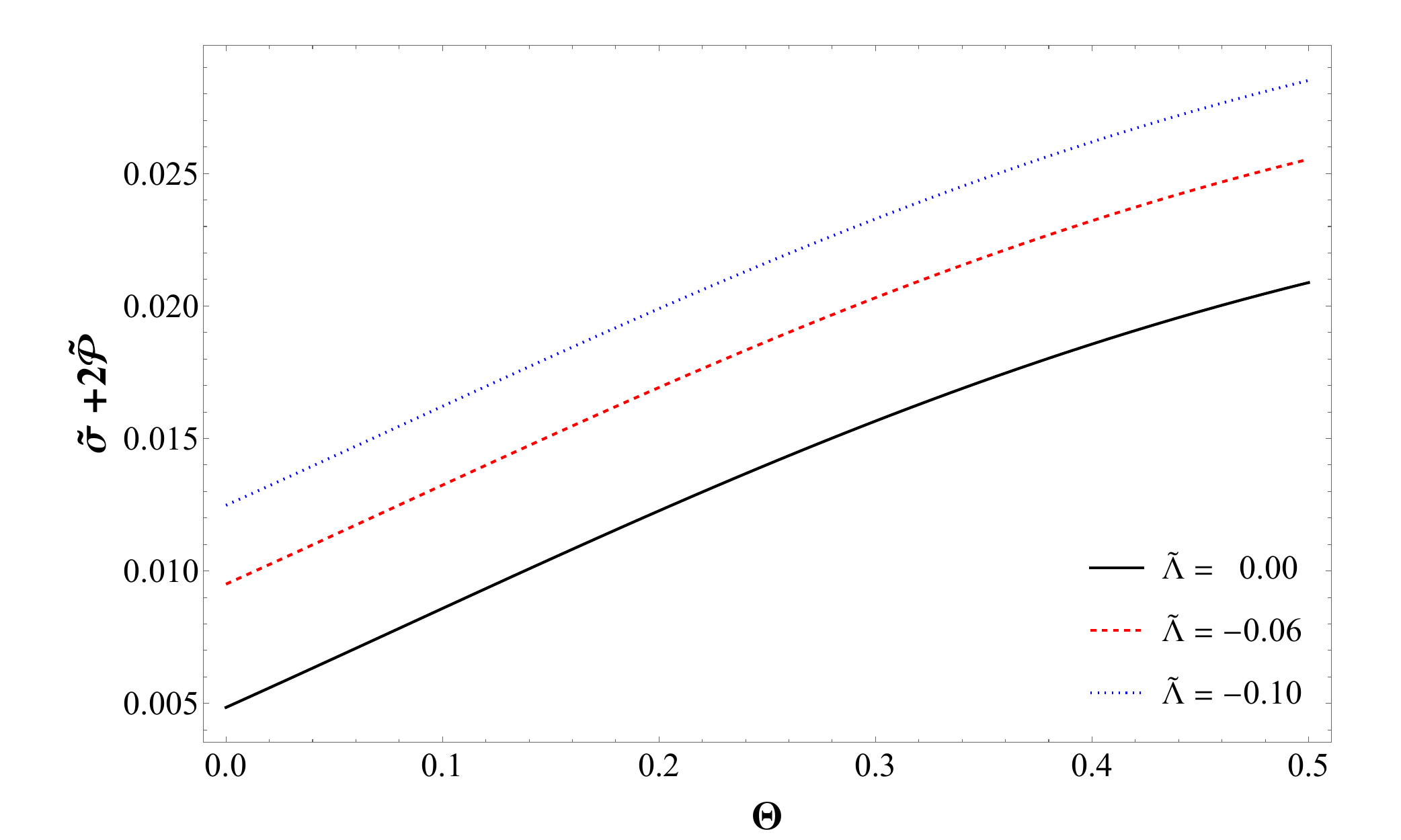}}
\subfigure[]{\includegraphics[scale=0.17]{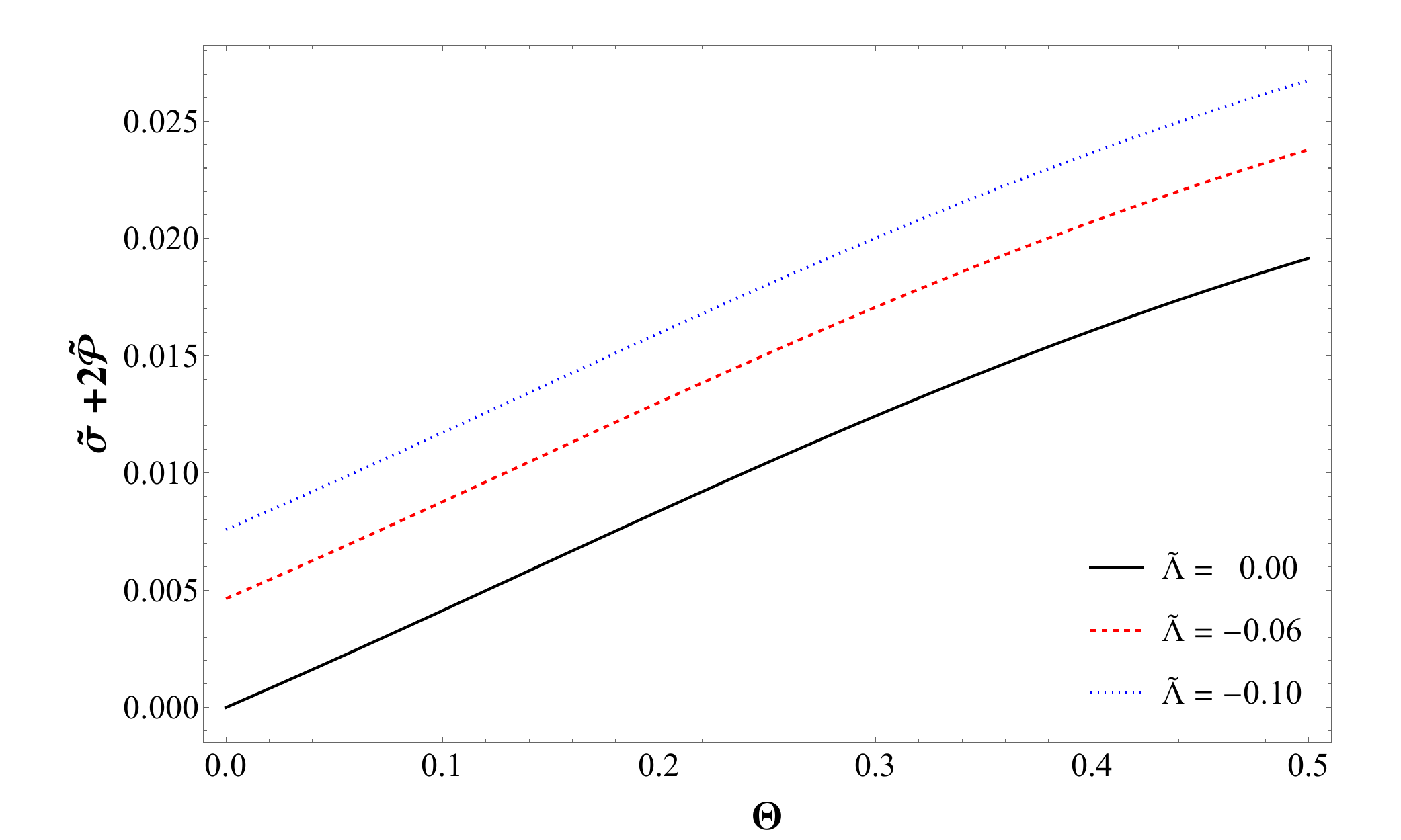}}
\caption{Quantities $\Tilde{\sigma} + 2\Tilde{\mathcal{P}}$ as a function of the noncommutative parameter \(\Theta\) for $\tilde{\Lambda} = 0$ (black, solid), $\Tilde{\Lambda} = -0.06$ (red, dashed) and $\Tilde{\Lambda} = -0.10$ (blue, dotted). The dimensionless Gauss-Bonnet coupling is (a) $\gamma = 0.03$ and (b) $\gamma = 0$.}
\label{SPres1}
\end{figure}

\begin{figure}[htbh!]
\centering
\subfigure[]{\includegraphics[scale=0.17]{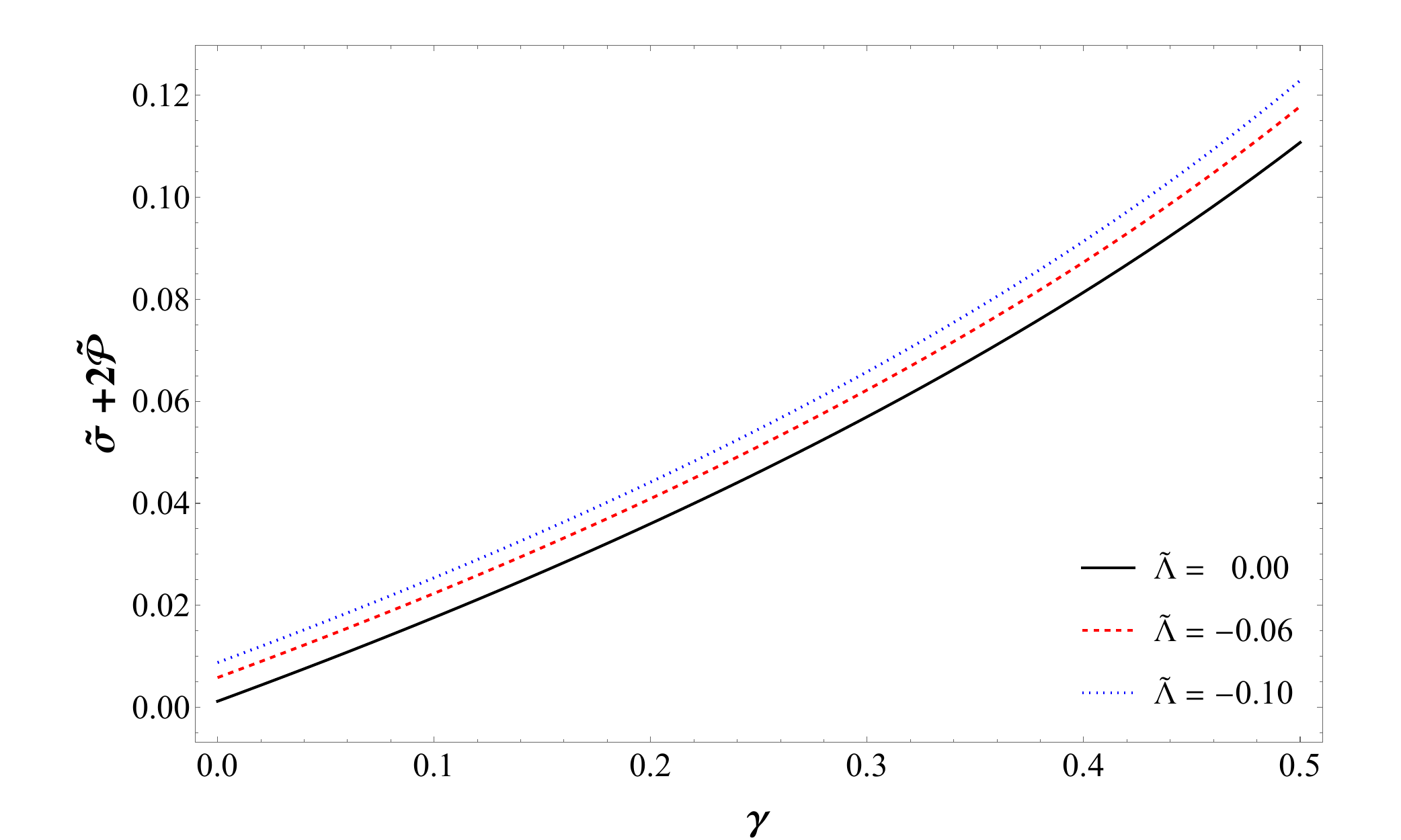}}
\subfigure[]{\includegraphics[scale=0.17]{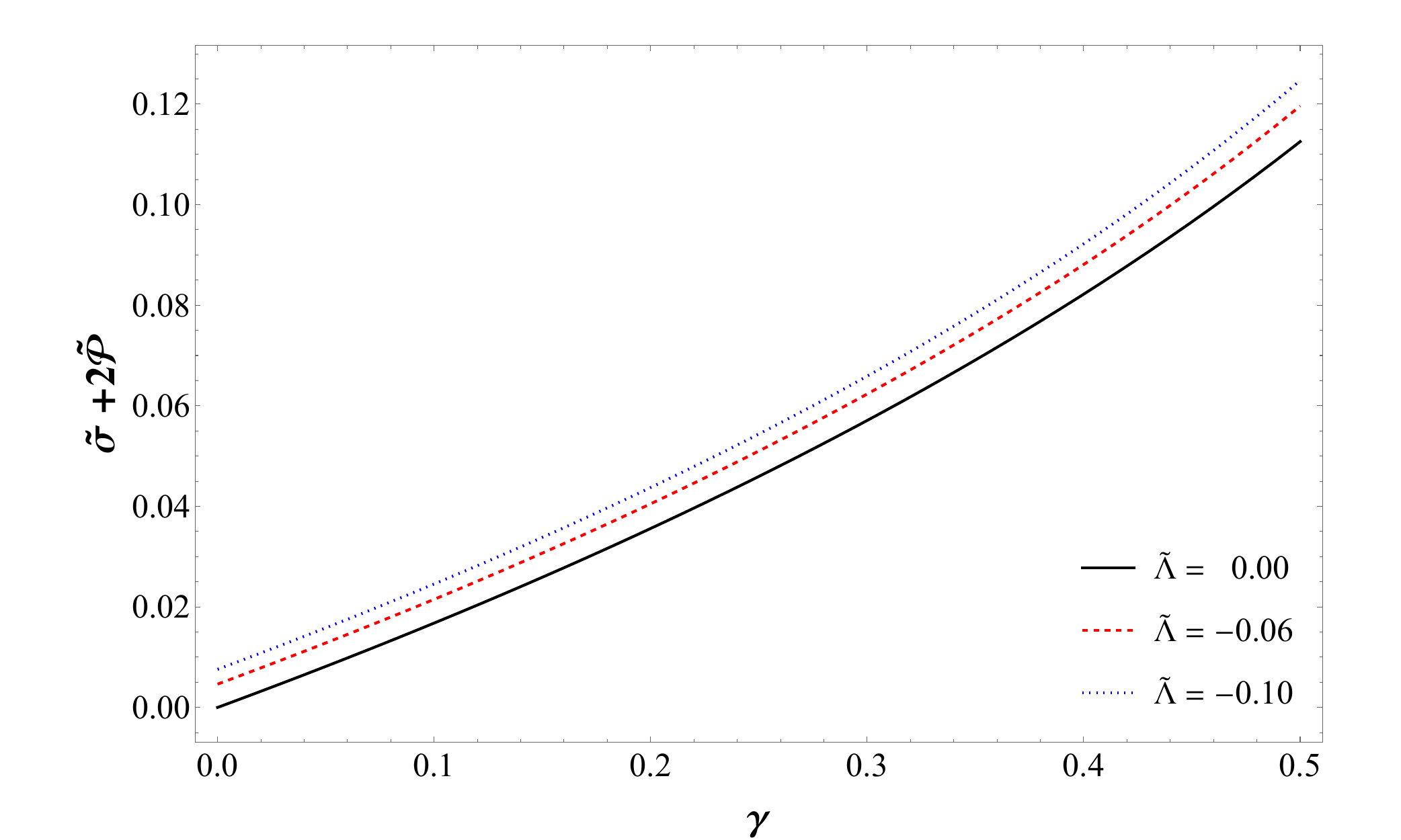}}
\caption{Quantities $\Tilde{\sigma} + 2\Tilde{\mathcal{P}}$ against the parameter $\gamma$ for  $\Tilde{\Lambda} = 0$ (black, solid), $\Tilde{\Lambda} = -0.06$ (red, dashed) and $\Tilde{\Lambda} = -0.10$ (blue, dotted). (a) are considered the noncommutativity parameter $\Theta=0.03$. (b) are considered the noncommutativity parameter $\Theta=0$.}
\label{SPres2}
\end{figure}

According to the energy conditions, the weak energy condition (WEC) must exist if $\sigma \geq 0$ and $\sigma + \mathcal{P} \geq 0$.  The null energy condition (NEC) is valid for continuity because $\sigma + \mathcal{P}\geq 0$.  To verify the strong energy condition (SEC), both $\sigma + \mathcal{P} \geq 0$ and $\sigma + 2 \mathcal{P} \geq 0$~\cite{Horvat:2007qa} must be satisfied.  Our calculations indicate that the pressure $\mathcal{P}$ (Figs. \ref{Pres1} and \ref{Pres2}) is positive, while the energy density $\sigma$ assumes negative values (Figs. \ref{Energy1} and \ref{Energy2}), implying a violation of the weak energy condition (WEC). In the context of thin-shell gravastars, the presence of such exotic matter at the interface is a well-known geometric requirement: it provides the extreme surface tension necessary to sustain the abrupt structural matching between the inner dark energy core characterized by the equation of state $ p =-\rho$ and the exterior spacetime. Thus, rather than being a direct property of the core itself, this WEC violation on the shell reflects the severe mechanical conditions required at the boundary to balance the repulsive interior, thereby preventing complete gravitational collapse and avoiding a central singularity. For sufficiently small $\Theta$, the positive pressure and energy density maintain $\tilde{\sigma}+2\tilde{\mathcal{P}} \geq 0$ (Figs. \ref{SPres1} and \ref{SPres2} ) and $\tilde{\sigma}+\tilde{\mathcal{P}} \geq 0$ which we can easily verify through the equations (\ref{sigadm}) and (\ref{presadm}) for sufficiently small values of $\tilde{\sigma}$ and $\Theta$ .

An isolated analysis of the Gauss–Bonnet coupling reveals that, despite meeting the minimum criteria necessary to satisfy the energy conditions, its isolated contribution does not induce the evolution of the energy density $\sigma$ (Fig.~\ref{Energy2}) or the pressure  $\mathcal{P}$ (Fig.~\ref{Pres2}) when $\Tilde{\Lambda}$ and $\Theta$ are set to zero.

We will now examine the scenario in which $\Lambda=0$, in which case the energy density and pressure equations above become
\FloatBarrier
\begin{eqnarray}
    \tilde{\sigma} = - \dfrac{1}{4 \pi } 
\left[ \sqrt{M - \gamma M^{2}} - \sqrt{M\left( 1- \frac{\Theta}{\sqrt{1+\Theta^{2}}}\right)-\gamma M^{2}\left( 1- \frac{2\Theta}{\sqrt{1+\Theta^{2}}}\right) } \, \right],
\label{sigadm}
\end{eqnarray}
\begin{eqnarray}
\tilde{\mathcal{P}} &=& \frac{1}{8 \pi } \left[\frac{M }{\sqrt{M - \gamma M^{2}}} -\dfrac{ M \left(1 - \dfrac{\Theta (1+ 2\Theta^{2})}{2(1+ \Theta^{2})^{3/2}}\right) - \dfrac{M^{2}\gamma \Theta}{(1 + \Theta^{2})^{3/2}} }{ \sqrt{ M\left( 1- \dfrac{\Theta}{\sqrt{1+\Theta^{2}}}\right)-\gamma M^{2}\left( 1- \dfrac{2\Theta}{\sqrt{1 +\Theta^{2}}}\right)} } \right],
\label{presadm}
\end{eqnarray}
\begin{eqnarray}
\tilde{\sigma}+2\tilde{\mathcal{P}} = \frac{1}{4\pi}\left[ 
\frac{M^{2}\gamma }{\sqrt{M  - \gamma M^{2}}}  + \frac{ M \left(\dfrac{2M\gamma \sqrt{1 +\Theta^{2}} }{2(1 +\Theta^{2})^{3/2}}+ P_{\Theta}\right)}{ \sqrt{ M\left( 1- \dfrac{\Theta}{\sqrt{1+\Theta^{2}}}\right)-\gamma M^{2}\left( 1- \dfrac{2\Theta}{\sqrt{1+\Theta^{2}}}\right)} } \right],
\label{s2Ptil-l}
\end{eqnarray}
where
\begin{eqnarray}
    P_{\Theta} = \dfrac{\left[\Theta - 6M \gamma \Theta + 2M\gamma\Theta^{2} \left(\sqrt{1+\Theta^{2}}-2\Theta \right) \right]}{2\left(1 +\Theta^{2}\right)^{3/2}}.
\end{eqnarray}
Thus, by examining the above result, we confirm that the existence of the minimum length preserves the criteria $\Tilde{\sigma}<0$ and $\Tilde{\mathcal{P}}>0$.

{Considering the limit $\Lambda=\Theta=0$, we observe that both the surface energy density, $\sigma$, and the surface pressure, $\mathcal{P}$, tend to zero, as illustrated in Figs.~(\ref{Energy2}) and (\ref{Pres2}). In this limit, the Gauss-Bonnet contribution alone yields vanishing surface stresses in the present gravastar configuration. This behavior indicates that the stability of the system relies on the combined effects of the cosmological constant and/or noncommutativity. Furthermore, when $\Lambda=\gamma=0$, the surface quantities $\tilde{\sigma}$ and $\tilde{\mathcal{P}}$ recover the same behavior reported in Ref.~\cite{Silva:2023bsu}. This result highlights the consistency of the present model, given that, in the absence of contributions associated with the cosmological constant and the Gauss-Bonnet coupling, the expressions obtained reduce to the limit already known in the literature.}

\subsection{Pure Gauss-Bonnet Contribution in the Exterior Region}\label{subsec:gb-exterior}
In this subsection, we analyze the case where the exterior region of the gravastar is described exclusively by the contribution of the Gauss-Bonnet coupling. This hypothesis allows us to investigate, in isolation, the effects of higher-order geometric corrections on the physical quantities of the thin layer and on stability of the configuration.

Using the general expressions derived previously and imposing that the exterior region be described exclusively by the contribution of the Gauss-Bonnet term, we obtain the following expressions for the surface energy density and surface pressure of the shell.
\begin{eqnarray}
    \tilde{\sigma} = - \dfrac{1}{4 \pi } 
\left[ \sqrt{M -\tilde{\Lambda} - \gamma (M-\tilde{\Lambda})^{2}} - \sqrt{M\left( 1- \frac{\Theta}{\sqrt{1+\Theta^{2}}}\right) } \, \right],
\label{sigadm}
\end{eqnarray}
\begin{eqnarray}
\Tilde{\mathcal{P}} &=& \frac{1}{8 \pi } \left[\frac{M -2 \Tilde{\Lambda} + 2\gamma\Tilde{\Lambda}^{2} - 2M\gamma \Tilde{\Lambda}}{\sqrt{M -\Tilde{\Lambda} - \gamma (M-\Tilde{\Lambda})^{2}}} -\dfrac{ M \left(1 - \dfrac{\Theta (1+ 2\Theta^{2})}{2(1+ \Theta^{2})^{3/2}}\right)}{ \sqrt{ M\left( 1- \dfrac{\Theta}{\sqrt{1+\Theta^{2}}}\right)} } \right],
\label{presadm}
\end{eqnarray}
\begin{eqnarray}
\Tilde{\sigma}+2\Tilde{\mathcal{P}} = \frac{1}{4\pi}\left[ 
\frac{M^{2}\gamma- \Tilde{\Lambda} -\Tilde{\Lambda}^{2} \gamma}{\sqrt{M -\Tilde{\Lambda} - \gamma (M-\Tilde{\Lambda})^{2}}}  + \frac{ \dfrac{M~\Theta }{2\left(1 +\Theta^{2}\right)^{3/2}}}{ \sqrt{ M\left( 1- \dfrac{\Theta}{\sqrt{1+\Theta^{2}}}\right)} } \right].
\label{s2Ptild}
\end{eqnarray}
\begin{figure}[htbh!]
\centering
\subfigure[]{\includegraphics[width=0.33\linewidth]{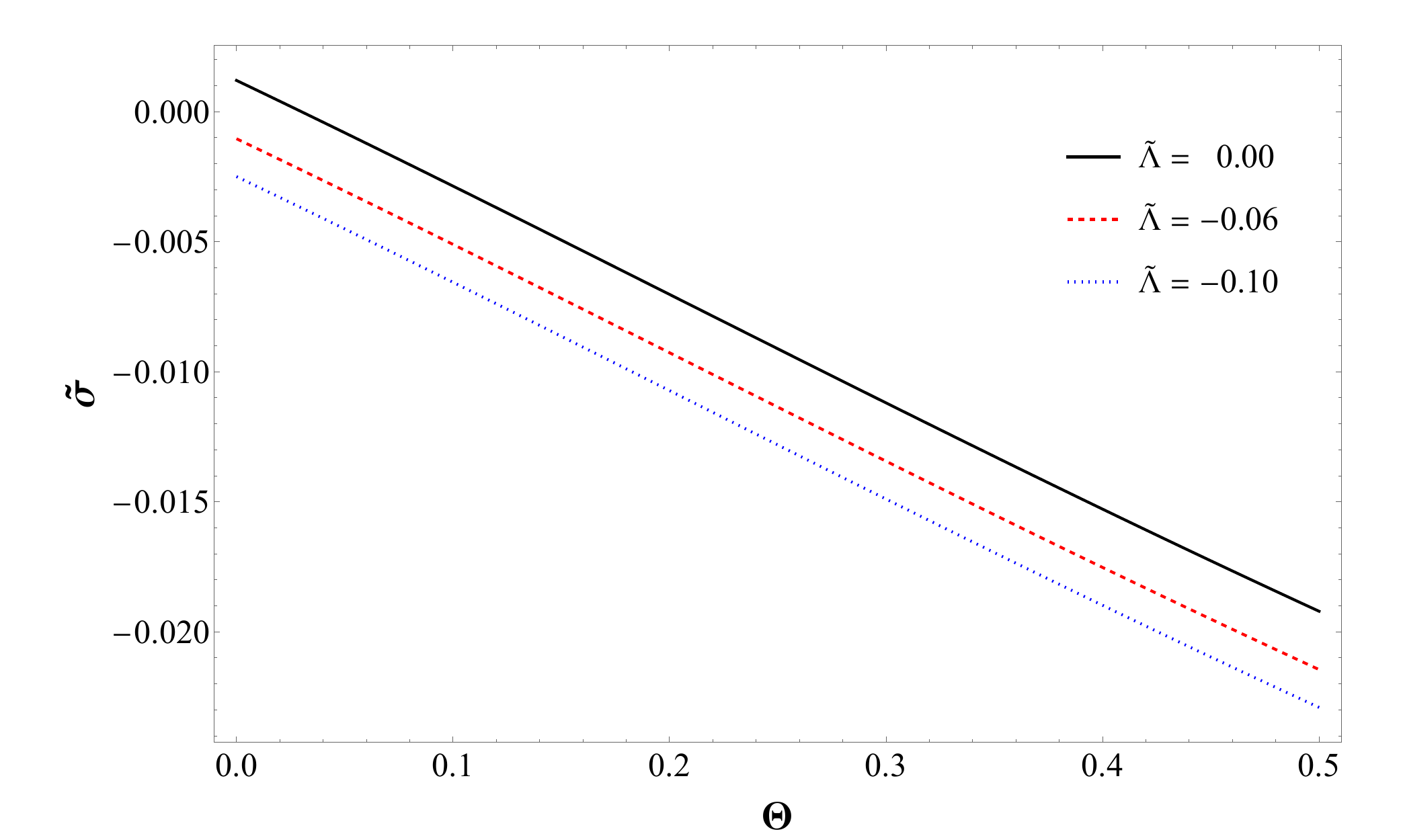}}\hfill%
\subfigure[]{\includegraphics[width=0.33\linewidth]{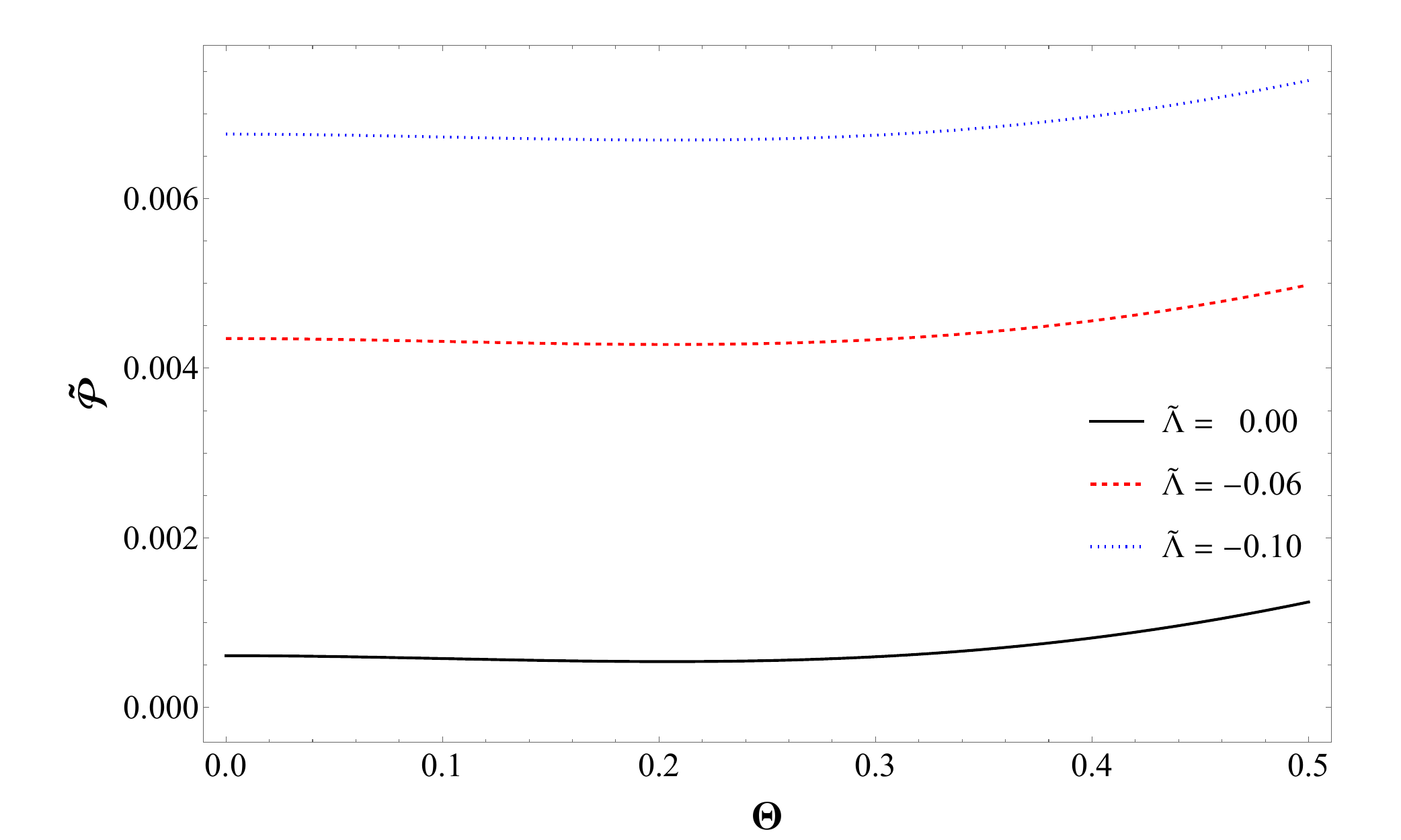}}\hfill%
\subfigure[]{\includegraphics[width=0.33\linewidth]{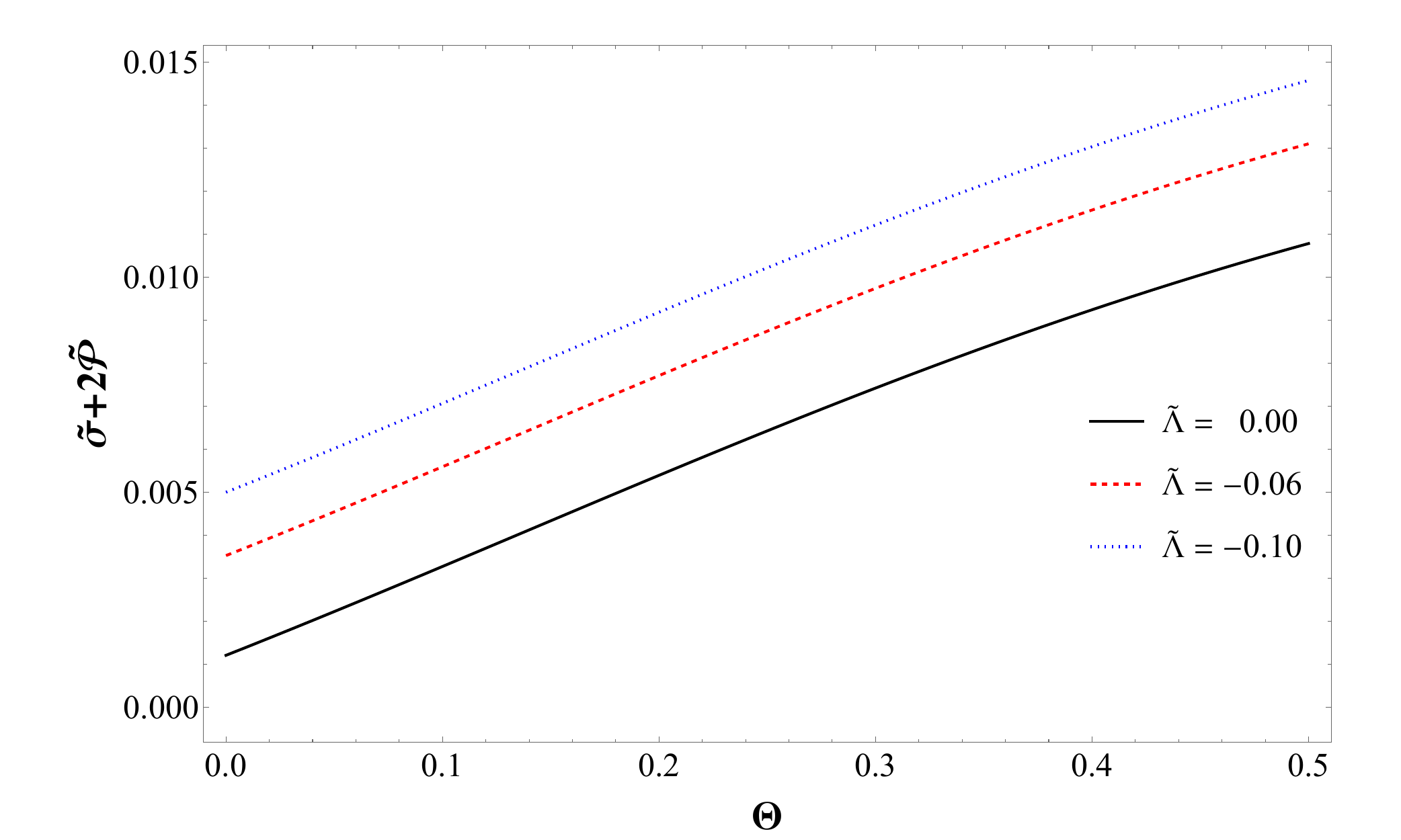}}
\caption{Shell quantities as functions of the noncommutative parameter \(\Theta\), for \(\tilde{\Lambda}=0\) (black solid), \(\tilde{\Lambda}=-0.06\) (red dashed), and \(\tilde{\Lambda}=-0.10\) (blue dotted), with \(M=1\) and \(\gamma=0.03\). (a) Surface energy density \(\tilde{\sigma}\). (b) surface pressure \(\tilde{\mathcal P}\). (c) combination \(\tilde{\sigma}+2\tilde{\mathcal P}\).}
\label{GB_ext}
\end{figure}

When the Gauss-Bonnet contribution is preserved solely in the exterior, Fig.(\ref{GB_ext}), and $\gamma=0.03$ is kept constant, the increase in $\Theta$ exclusively affects the noncommutative inner region. In this scenario, the surface density becomes increasingly negative with $\Theta$. For $\tilde{\Lambda}=0$, the configuration initially exhibits a small positive density at $\Theta=0$; however, $\tilde{\sigma}$ crosses zero near $\Theta\simeq 0.03$ and remains negative for higher values. For $\tilde{\Lambda}=-0.06$ and $-0.10$, the density is already negative at $\Theta=0$ and further decreases as the noncommutativity intensifies. Consequently, while holding the strength of the exterior Gauss-Bonnet coupling constant, it is the enhanced noncommutative deformation of the interior that demands a more negative surface density at the junction.

Conversely, the pressure remains positive throughout the analyzed interval. Its dependence on $\Theta$ is significantly weaker than that of $\tilde{\sigma}$, exhibiting only a slight initial decrease followed by a subsequent rise. Thus, the predominant effect of increasing $\Theta$ is not a change in the sign of the pressure, but rather an increase in the magnitude of the negative density. Although the pressure remains positive, it gradually ceases to offset the negative contribution of $\tilde{\sigma}$.

For $\tilde{\Lambda}=0$, the WEC can only be met within a narrow region prior to the sign reversal of $\tilde{\sigma}$, approximately for $\Theta \lesssim 0.03$. Beyond this point, even though the NEC remains satisfied over a short interval, the WEC is already violated, given that $\tilde{\sigma} < 0$. For $\tilde{\Lambda}=-0.06$ and $-0.10$, the WEC remains violated throughout the entire analyzed range, as the surface density never assumes positive values.
\subsection{Pure Gauss-Bonnet Contribution in the Interior Region}\label{subsec:gb-interior}
In this subsection, we consider the case where only the interior region of the gravastar is modified by the contribution of the Gauss-Bonnet coupling, while the exterior geometry remains described by the standard solution.
\begin{eqnarray}
    \Tilde{\sigma} = - \dfrac{1}{4 \pi } 
\left[ \sqrt{M -\Tilde{\Lambda}} - \sqrt{M\left( 1- \frac{\Theta}{\sqrt{1+\Theta^{2}}}\right)-\gamma M^{2}\left( 1- \frac{2\Theta}{\sqrt{1+\Theta^{2}}}\right) } \, \right],
\label{siga_in}
\end{eqnarray}
\begin{eqnarray}
\Tilde{\mathcal{P}} &=& \frac{1}{8 \pi } \left[\frac{M -2 \Tilde{\Lambda} }{\sqrt{M -\Tilde{\Lambda}}} -\dfrac{ M \left(1 - \dfrac{\Theta (1+ 2\Theta^{2})}{2(1+ \Theta^{2})^{3/2}}\right) - \dfrac{M^{2}\gamma \Theta}{(1 + \Theta^{2})^{3/2}} }{ \sqrt{ M\left( 1- \dfrac{\Theta}{\sqrt{1+\Theta^{2}}}\right)-\gamma M^{2}\left( 1- \dfrac{2\Theta}{\sqrt{1 +\Theta^{2}}}\right)} } \right],
\label{pres_in}
\end{eqnarray}
\begin{eqnarray}
\Tilde{\sigma}+2\Tilde{\mathcal{P}} = \frac{1}{4\pi}\left[ 
\frac{ -\Tilde{\Lambda}}{\sqrt{M -\Tilde{\Lambda}}}  + \frac{ M \left(\dfrac{2M\gamma \sqrt{1 +\Theta^{2}} }{2(1 +\Theta^{2})^{3/2}}+ P_{\Theta}\right)}{ \sqrt{ M\left( 1- \dfrac{\Theta}{\sqrt{1+\Theta^{2}}}\right)-\gamma M^{2}\left( 1- \dfrac{2\Theta}{\sqrt{1+\Theta^{2}}}\right)} } \right],
\label{s2P_in}
\end{eqnarray}
where
\begin{eqnarray}
    P_{\Theta} = \dfrac{\left[\Theta - 6M \gamma \Theta + 2M\gamma\Theta^{2} \left(\sqrt{1+\Theta^{2}}-2\Theta \right) \right]}{2\left(1 +\Theta^{2}\right)^{3/2}}.
\end{eqnarray}
\begin{figure}[htbh!]
\centering
\subfigure[]{\includegraphics[width=0.33\linewidth]{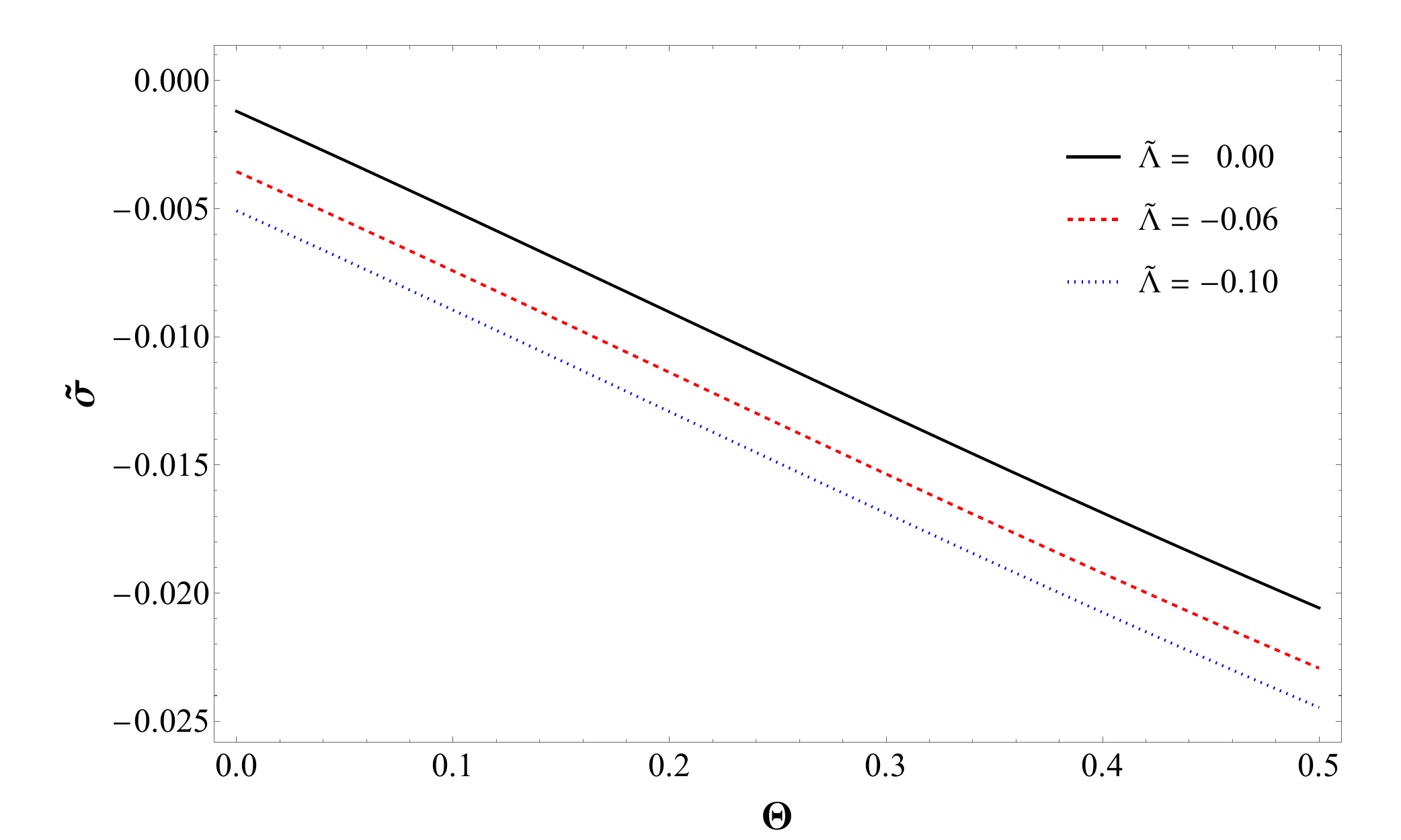}}\hfill%
\subfigure[]{\includegraphics[width=0.33\linewidth]{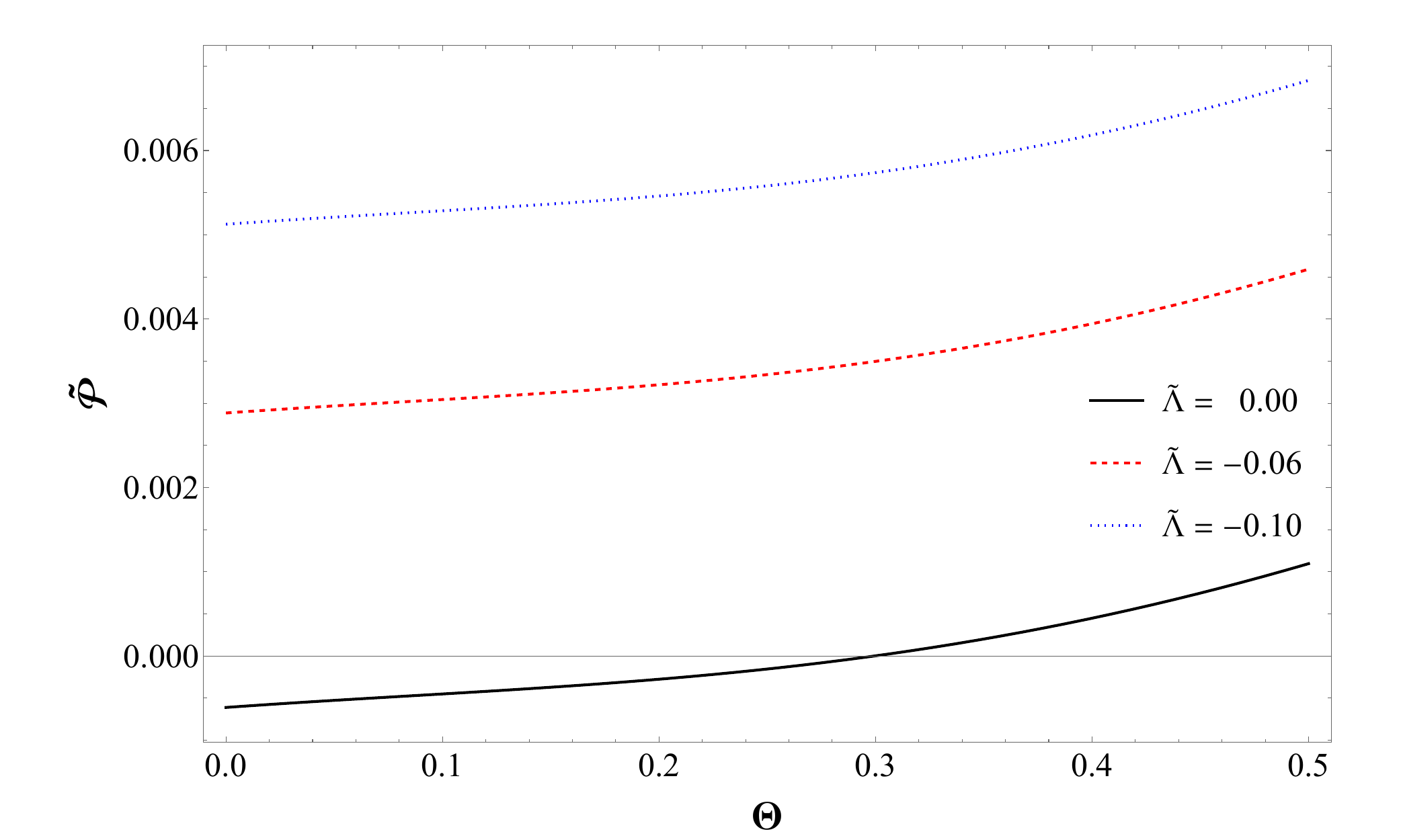}}\hfill%
\subfigure[]{\includegraphics[width=0.33\linewidth]{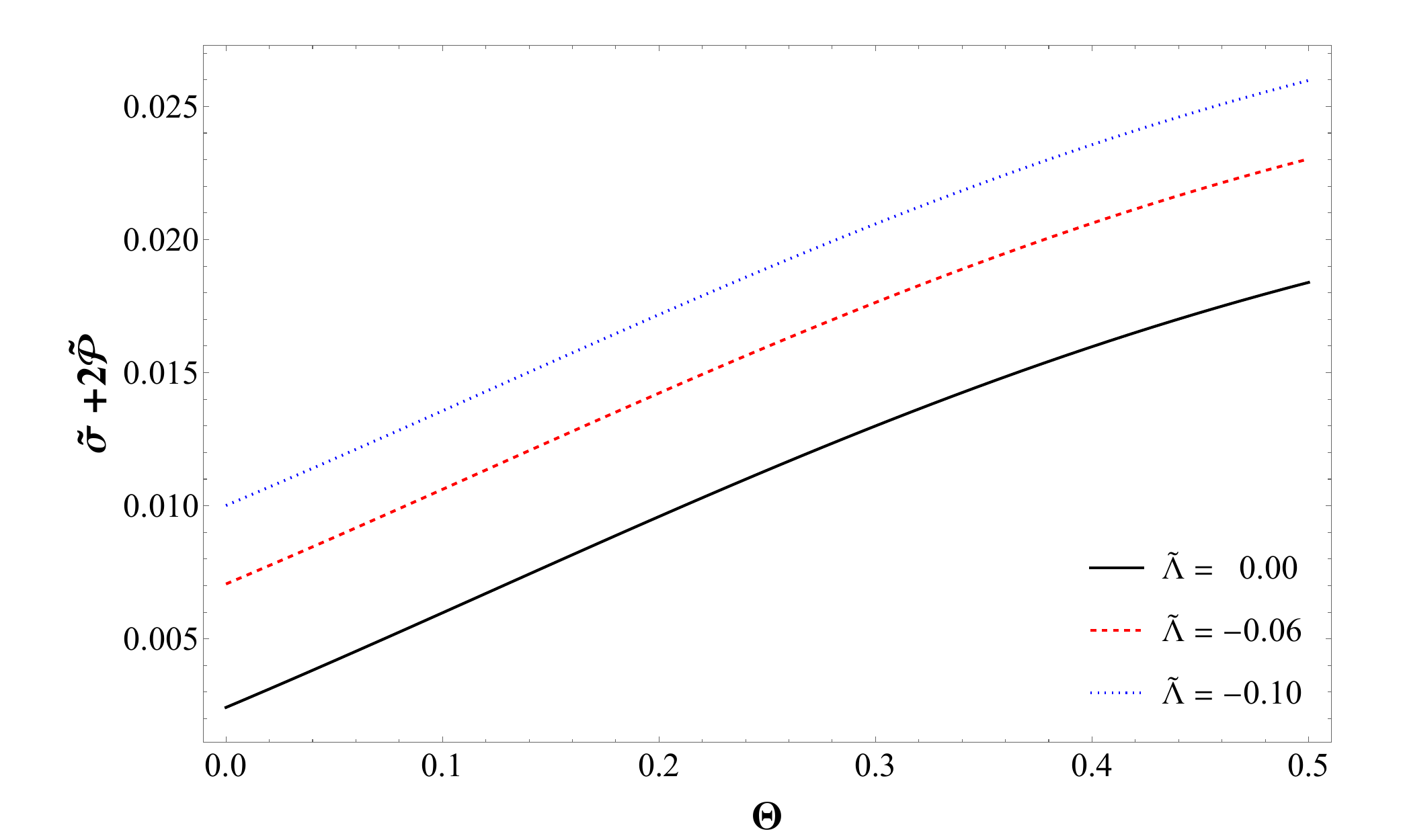}}
\caption{Influence of the noncommutative parameter \(\Theta\) on the matter content of the shell when the Gauss-Bonnet contribution is restricted to the interior geometry, for \(M=1\) and \(\gamma=0.03\). The curves correspond to \(\tilde{\Lambda}=0\) (black solid), \(\tilde{\Lambda}=-0.06\) (red dashed), and \(\tilde{\Lambda}=-0.10\) (blue dotted). The surface energy density in panel (a) remains negative and decreases as \(\Theta\) increases. In panel (b), the pressure grows with \(\Theta\), with the \(\tilde{\Lambda}=0\) branch changing sign within the displayed interval. Panel (c) shows that \(\tilde{\sigma}+2\tilde{\mathcal P}\) remains positive and increases throughout the range considered.}
\label{GB_in}
\end{figure}
\FloatBarrier
When the Gauss-Bonnet correction is confined solely to the interior, Fig.(\ref{GB_in}), again maintaining a fixed $\gamma=0.03$, increasing $\Theta$ exerts an even more direct effect on the shell matter, as both deformations now operate within the same geometric sector. The surface density is negative from $\Theta=0$ onward for all considered values of $\tilde{\Lambda}$, becoming increasingly negative as $\Theta$ grows. Consequently, in this scenario, there is no transition to a positive density within the analyzed range. Eqs. (\ref{siga_in})–(\ref{s2P_in}) themselves illustrate that $\gamma$ and $\Theta$ appear jointly in the interior contribution.

For $\tilde{\Lambda}=-0.06$ and $-0.10$, the pressure remains positive and increases with $\Theta$. For $\tilde{\Lambda}=0$, however, it begins at a small negative value at $\Theta=0$ and undergoes a sign reversal only around $\Theta \simeq 0.30$. Therefore, even when the pressure increases with noncommutativity, this growth is insufficient to offset the increasing magnitude of the negative surface density.

This distinction becomes apparent in the energy conditions. Since $\tilde{\sigma}<0$ throughout the entire interval, the WEC remains violated for all three choices of $\tilde{\Lambda}$. The NEC is also significantly more restrictive than in the case where the Gauss-Bonnet correction operates solely in the exterior. For $\tilde{\Lambda}=0$ and $-0.06$, the combination $\tilde{\sigma}+\tilde{\mathcal P}$ remains negative across the entire range $0\leq\Theta\leq0.5$. For $\tilde{\Lambda}=-0.10$, it is only slightly positive very close to $\Theta=0$ and becomes negative almost immediately, around $\Theta\simeq 0.001$.

Thus, for a fixed $\gamma$, enhancing the noncommutativity in the interior sector does not lead to the restoration of the energy conditions. On the contrary, the surface density becomes increasingly negative, and the positive pressure, when present, does not grow sufficiently to maintain $\tilde{\sigma}+\tilde{\mathcal P}\geq0$. This explains why the NEC remains violated across virtually the entire studied range of $\Theta$. Also in this scenario, the quantity $\tilde{\sigma}+2\tilde{\mathcal P}$ remains positive throughout the analyzed interval. Therefore, the critical constraint for the strong energy condition (SEC) arises once again from $\tilde{\sigma}+\tilde{\mathcal P}$.

\section{Conclusions}\label{con}

We have constructed a $(2+1)$-dimensional thin-shell gravastar by matching a noncommutative BTZ interior incorporating Gauss--Bonnet corrections to a Gauss--Bonnet-modified BTZ exterior. Using the junction formalism, we derived the surface energy density and pressure of the shell and investigated how the noncommutative parameter, the Gauss--Bonnet coupling, and the cosmological constant modify the surface stress-energy distribution.

The thermodynamic analysis shows that both the noncommutative and Gauss--Bonnet corrections become particularly relevant in the small-horizon regime. The corrected Hawking temperature develops a finite maximum at a nonzero horizon radius, while the heat capacity exhibits a positive branch associated with a locally thermodynamically stable regime. These results show that the deformation parameters modify the thermal structure of the BTZ geometry predominantly at small scales. In the limit in which the deformation parameters vanish, the corresponding BTZ results are recovered.

For the thin shell, the configurations investigated here are predominantly characterized by a negative surface energy density and a positive surface pressure over the parameter ranges considered. The negative value of $\tilde{\sigma}$ implies a violation of the weak energy condition at the junction, indicating that exotic surface stresses are required to connect the interior and exterior geometries.

The null and strong energy conditions exhibit a parameter-dependent behavior governed by the balance between the surface energy density and pressure. When the Gauss--Bonnet correction is restricted to the exterior region, these conditions can be satisfied within limited parameter intervals, whereas the interior-only correction leads to a more restrictive behavior, with the null energy condition violated over almost the entire analyzed range.

An important feature of the junction analysis is that the effect of the Gauss--Bonnet correction depends on the geometric sector in which it is introduced. When the correction is restricted to the exterior region, increasing the Gauss--Bonnet coupling tends to shift the surface energy density toward less negative values within the parameter range considered. When the correction is restricted to the interior region, the opposite tendency is observed. In the latter case, the Gauss--Bonnet and noncommutative deformations act on the same metric sector, giving rise to explicit mixed contributions involving $\gamma$ and $\Theta$ in the surface stresses.

This comparison also clarifies the limit $\tilde{\Lambda}=\Theta=0$. Within the present construction, when the same Gauss--Bonnet deformation is included on both sides of the junction, the corresponding contributions to the surface stresses cancel in this limit, yielding $\tilde{\sigma}=\tilde{\mathcal P}=0$. Thus, the shell stresses are governed by the discontinuity between the two geometries rather than by the Gauss--Bonnet coupling alone. When the correction is introduced only on one side of the junction, this cancellation no longer occurs and nonvanishing surface stresses are obtained.

Finally, in the limit $\tilde{\Lambda}=\gamma=0$, the expressions for the surface quantities reduce to the previously studied noncommutative BTZ gravastar case, providing a consistency check of the present construction. The model therefore provides a framework for disentangling the effects of noncommutative smearing, higher-curvature corrections, and their location with respect to the junction surface in lower-dimensional compact configurations.

 \begin{acknowledgments}
We thank CNPq and CAPES for partial financial support. M.A.A  acknowledges support from CNPq (Grant no. $301683/2025-5$). A.T.N.S thanks the Paraíba State Research Support Foundation (FAPESQ) (Grant No. $765/2026$) for financial support. L.C. acknowledges support from CNPq (Grant no. $306842/2026-2$).
\end{acknowledgments}

\end{document}